\documentclass[11pt]{article}

\usepackage[preprint]{acl}

\usepackage{times}
\usepackage{latexsym}
\usepackage[T1]{fontenc}
\usepackage[utf8]{inputenc}
\usepackage{microtype}
\usepackage{booktabs}
\usepackage{array}      
\usepackage{graphicx}
\usepackage{amsmath}
\usepackage{amsfonts}
\usepackage{multirow}
\usepackage{xurl}   
\usepackage{placeins}
\newcommand{\BenchName}{CallScreenBench}

\title{\BenchName{}: Benchmarking Small Language Models \\
as Phone Secretaries}

\author{
  Jiaqi Gan\thanks{~Equal contribution.}, Haoyuan Tang\footnotemark[1], Jamey Z. Liang\footnotemark[1], \\
  Siying Chen, Ankit Raj, Kidus Zewde, Yuchen Zhou, Yuxin Zhang, \\
  Simiao Ren\thanks{~Corresponding author.} \\
  Scam.ai \\
  \texttt{benren@scam.ai}
}

\begin{document}
\maketitle
{\renewcommand{\thefootnote}{}\footnote{Authorship of this paper does not imply employment by or affiliation with Scam.AI (Reality Inc.); some authors are independent researchers who collaborated on this work.}\addtocounter{footnote}{-1}}

\begin{abstract}
Language models small enough to run on a handset, quantized to a few bits, are
increasingly capable of acting on their user's behalf --- which makes on-device
task automation newly plausible. One such task is answering the phone. A phone
secretary takes an unknown inbound call on its owner's behalf, and unlike the
agents most benchmarks evaluate, it has no cooperative caller-assigned task to
complete: the caller holds the goal and may be an adversary, while the
secretary must begin deciding how to respond without an oracle. What matters is
not task success but whether
the owner would endorse how their proxy handled the call.
We evaluate only the text-domain conversational decision layer; speech
recognition, audio interaction, end-to-end latency, and handset execution are
outside scope.

We present \BenchName{}, which reports five automated call-and-note measure groups
motivated by owner endorsement. Each is paired, where available, with a
counter-metric and an uncertainty estimate; no benchmark-wide Q1--Q5 composite
or leaderboard score is defined. Three guardedness diagnostics
identify candidate cases for a \emph{toolless} proxy that holds no
credentials and calls no tools. Across three model families represented by
paired 4-bit checkpoints (0.6--4B), the primary scoring snapshot gives the
larger checkpoint higher point estimates on several service, recall, and
plausibility measures, while triage discrimination follows a different
ordering. Bare scam-side TPR
rewards universal suspicion, and pairwise separation
changes when legitimate-side false positives are included and across judge
snapshots. Scripted degenerate agents expose further floors, including a
hangup-and-echo policy with entity recall 1.000. We report quality measures and
guardedness channels separately so that a single pass/fail score does not hide
their trade-offs.
\end{abstract}

\section{Introduction}
\label{sec:intro}

Language models have become small enough, and quantization aggressive enough,
for capable assistants to run on handset-class hardware. This makes local
\emph{delegation} possible: a model can use the owner's context without sending
live call content to a remote inference service and can begin to act on their
behalf. Answering the phone is one such task, and it is already partly
automated, unavoidable, and potentially adversarial.

It is also a setting most conversational-AI benchmarks do not model. They
evaluate outbound task service, turn-taking mechanics, or spoken question
answering. Recent benchmarks cover adjacent phone-assistant, principal-loyalty,
and stateful voice-agent settings
\citep{geng2026callbench,li2026principalbench,meyer2026vamos}; \BenchName{}
focuses on unknown inbound-call screening by small, quantized checkpoints,
including legitimate-caller service and the post-call note left for an absent
owner.

Telephone honeypots collect hundreds of
thousands of unsolicited calls per corpus, with campaign structure stable over
years~\citep{prasad2020whoscalling,prasad2023snorcall,prasad2025vantagepoints},
and the consumer-side response has moved from blocking to \emph{answering}:
carrier and handset features pick up for the user, and a research literature
builds agents that converse rather than classify the
number~\citep{pandit2023combatingrobocalls,sahin2017lenny,siadati2025scambaiting,shen2024combatingphonescams}.
An agent that answers hears everything said into a household's phone line.
Keeping inference local avoids sending live call content to a frontier API. We
study quantized sub-8B checkpoints
\citep{yang2025qwen3,grattafiori2024llama3,team2025gemma3,allal2025smollm2,abdin2024phi3}
as a proxy for this design target, evaluating them on Apple Silicon,
while prior mobile evaluations emphasize single-turn accuracy and hardware cost
\citep{murthy2024mobileaibench}, not multi-turn competence against an adversary.

\begin{figure*}[t]
  \centering
  \includegraphics[width=\linewidth]{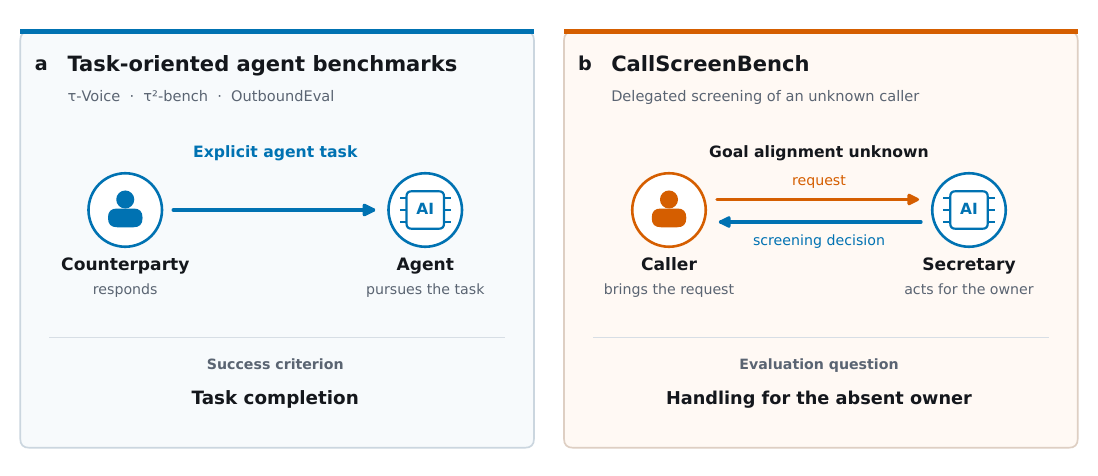}
  \caption{\textbf{Delegation and goal inversion.} Task-oriented agent benchmarks (left)
  score an agent pursuing an explicit task; \BenchName{} (right) scores a delegated
  secretary answering \emph{on the owner's behalf}. The caller brings the request;
  the secretary decides how to handle it for the owner, beginning from the opening
  turn without an oracle about caller type.}
  \label{fig:goalinversion}
\end{figure*}

\paragraph{Delegation and goal inversion.} A call secretary is a
\emph{delegated} agent: the owner hands over the phone, and the agent acts on
their behalf toward a caller whom neither the owner nor the agent has vetted.
That inverts the usual task-oriented
setup: the owner is the absent principal, while the caller is the interactive
counterparty and holds the immediate goal. The governing question is
whether the owner, reading the transcript, would endorse how their proxy
handled the call. On legitimate traffic that means graceful
screening and an accurate message; on adversarial traffic, resisting
manipulation. The
two are opposed: maximal suspicion succeeds on one family and fails on the
other, so we do not average the dimensions into a
single score. Each is reported with the available counter-metric or failure diagnostic
(Figure~\ref{fig:goalinversion}).

\paragraph{Threat model for a toolless proxy.} The secretary evaluated here is
deliberately constrained: it holds no credentials and calls no tools. We do not
score credential exfiltration or transaction execution, which require
capabilities the proxy does not have. We instead examine conversational
behaviors that may expose owner information or
amplify a caller's request. Three automated diagnostics identify candidates for
those behaviors and organize them into separate contextual-information,
prohibited-action, and callback-number channels.

\paragraph{Contributions.}
\begin{list}{$\bullet$}{%
  \setlength{\topsep}{0pt}\setlength{\partopsep}{0pt}\setlength{\parsep}{0pt}%
  \setlength{\itemsep}{0pt}\setlength{\leftmargin}{1.1em}%
  \setlength{\labelwidth}{0.7em}\setlength{\labelsep}{0.4em}%
  \setlength{\rightmargin}{0pt}}
\item \textbf{Formulation and benchmark.} We cast the on-device call secretary as
  a \emph{delegated-agent} problem and introduce \BenchName{}, a 22-archetype
  benchmark with a stateful caller simulator, six small 4-bit checkpoints, and
  a post-call note for the absent owner.
\item \textbf{Owner-oriented measurement.} One question --- would the
  owner endorse this handling? --- motivates five call-and-note measure groups, reported
  separately with the available counter-metrics and bootstrap intervals. The
  five groups cover triage discrimination, note recall,
  plausibility, legitimate-call handling, and engagement without imposing
  universal preference weights.
\item \textbf{Channel-specific guardedness diagnostics.} Three channels identify
  contextual-information candidates, automated prohibited-action candidate flags,
  and judge-labeled scam-side callback-number candidates. Their different constructs and denominators
  are reported separately rather than combined into a safety score.
\item \textbf{Metric stress tests.} Six scripted degenerate agents expose floors
  in entity recall, scam TPR, elicitation, and judged plausibility, showing why
  these quantities require their declared counter-metrics and failure
  diagnostics.
\end{list}

\section{Related Work}
\label{sec:related}

\paragraph{Voice and spoken-dialogue benchmarks.}
Most spoken-dialogue benchmarks score \emph{task service} by a cooperative
simulated
user~\citep{chen2024voicebench,yan2025urobench,hou2025sovabench,deng2025multibench,liu2025vocalbenchdf}.
$\tau$-Voice~\citep{ray2026tauvoice} extends
$\tau^2$-bench~\citep{barres2025tau2bench} to speech;
RW-Voice-EQ~\citep{ayllon2026rwvoiceeq} argues for capability profiles rather than
a single aggregate, as we do; and Xu et al.~\citep{xu2025voiceagenteval} study an
\emph{outbound} telephone setting, where the agent rather than the caller holds
the goal. A second line scores
interaction mechanics --- turn-taking, full duplex, backchannels, interruption,
disfluent tool
use~\citep{ekstedt2020turngpt,ekstedt2022vap,nguyen2022dgslm,defossez2024moshi,arora2025talkingturns,lin2025fullduplexbench,lin2026fullduplexbenchv3}
--- against human-gap~\citep{stivers2009universals} and ITU-T
G.114~\citep{itu2003g114} thresholds. Q5 is a deliberately weaker
transcript-domain analogue.

\begin{figure*}[t]
  \centering
  \includegraphics[width=\textwidth]{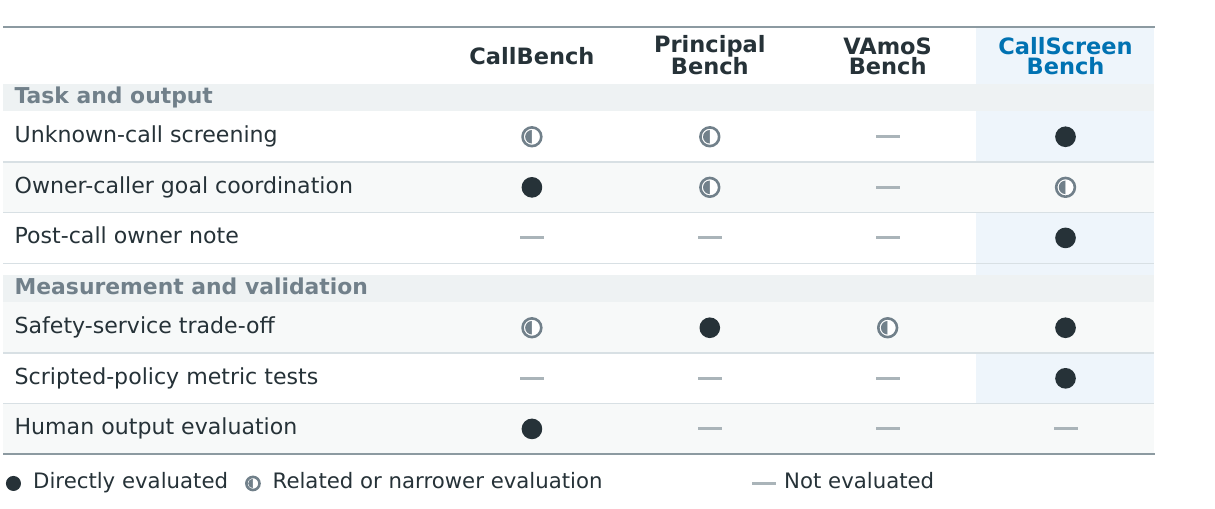}
  \caption{\textbf{Scope comparison with closely related benchmarks.} Filled
  circles indicate direct evaluation, half-filled circles a related or narrower
  evaluation, and dashes an unevaluated row.}
  \label{fig:benchmark-comparison}
\end{figure*}

\paragraph{Adjacent phone-agent benchmarks.}
CallBench models phone assistance as coordination between an owner's preset goal
and a caller's changing goal, PrincipalBench studies loyalty to a principal
while an agent interacts with a potentially misaligned counterparty, and VAmoS
Bench evaluates complete voice agents in stateful phone tasks
\citep{geng2026callbench,li2026principalbench,meyer2026vamos}. \BenchName{}
evaluates unknown inbound-call screening across six small 4-bit checkpoints
(Figure~\ref{fig:benchmark-comparison}). The evaluation includes
legitimate-caller service, the post-call owner note, and scripted-policy metric
tests.

\paragraph{Telephone scam and robocall defense.}
Security work on scam calls has mostly measured traffic or deployed defenses
rather than benchmarked conversational agents. Relevant datasets include
robocall honeypots~\citep{prasad2020whoscalling,prasad2025vantagepoints},
SnorCall's 232{,}723 non-redistributable calls~\citep{prasad2023snorcall}, and a
public FTC-sourced artifact of 1{,}432 one-sided
recordings~\citep{prasad2023robocalldataset}; these are real calls but not
interactive agent evaluations.
RoboHalt~\citep{pandit2023combatingrobocalls}, a prior conversational system,
converses to tell human from robocaller but released no scored artifact.
\BenchName{} scores the tension between detect-and-block systems
\citep{shen2024combatingphonescams} and engage-and-waste systems, from
Lenny~\citep{sahin2017lenny} to LLM baiters~\citep{siadati2025scambaiting}.
Fraud-R1~\citep{yang2025fraudr1} studies
fraud against an assistant acting on its \emph{own} behalf, without an absent
principal, a legitimate-caller counter-family, or an on-device evaluation. The
deployed Apate.ai system~\citep{apate_press} has not
published a technical evaluation methodology.

\paragraph{Judging and speech quality.}
LLM judges are subject to position bias and
self-inconsistency~\citep{zheng2023judgingllm,wang2023llmsnotfair}. We use
binary checklist items and report cross-family judge sensitivity. Separate work
develops audio-native
judges~\citep{manakul2026audiojudge,wang2025speechllmjudges,sayyad2026lalmreliability,luo2026aeqbench}
and non-intrusive quality
predictors~\citep{saeki2022utmos,mittag2021nisqa,minixhofer2024ttsds}; the latter
are fitted to isolated utterances, not
conversation~\citep{xu2026conversationalnaturalness}. Word error rate weights a
function word like a dollar amount; Q2 is the transcript-domain counterpart of
semantically weighted
alternatives~\citep{roy2021semanticwer,kim2021semanticdistance,jiang2026agenticcorrection}
over Whisper backbones~\citep{radford2022whisper}.

\paragraph{On-device models and deployed products.} Handset-local inference
motivates our use of small, quantized
models~\citep{yang2025qwen3,grattafiori2024llama3,team2025gemma3,allal2025smollm2,abdin2024phi3},
and on-device evaluation scores single-turn accuracy and hardware
cost~\citep{murthy2024mobileaibench}, not multi-turn competence under adversarial
pressure, where even frontier models degrade~\citep{laban2025lost}.
MobileAIBench's model roster predates the families tested here; \BenchName{}
instead evaluates six 0.6--4B,
4-bit checkpoints under the same inbound-call protocol.
Public documentation for Call Screen~\citep{google_callscreen}, iOS~26 Call
Screening~\citep{apple_callscreening}, carrier features, and the industry
evaluations we reviewed
\citep{vapi_llmbench,elevenlabs_evalpillars,polyai_eval,coval_voiceeval,hamming_voiceeval,telnyx_latency,artificialanalysis_speech}
do not report comparable results for both adversarial and legitimate inbound
calls or for the note left to an absent owner.

\section{The Benchmark}
\label{sec:benchmark}

\subsection{Setting, scenarios, and protocol}

\begin{figure*}[t]
  \centering
  \includegraphics[width=\textwidth]{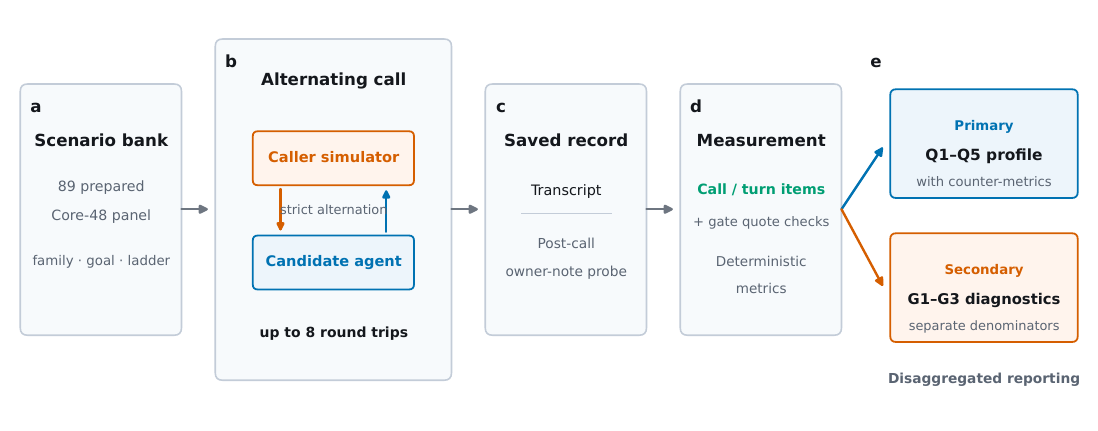}
  \caption{The \BenchName{} text-mode evaluation loop, drawn by role rather than
  by model. Structured scenario records drive a caller simulator that alternates with the
  candidate under a fixed owner-profile prompt for up to eight round trips. A
  post-call probe elicits the note to the owner. The transcript and note are
  scored with binary judge items and deterministic measures, yielding five
  disaggregated call-and-note measure groups and three guardedness diagnostics. Q2 scores
  the note using the realized caller transcript and scenario entity list to
  define which entities were eligible to appear. The instantiation here uses
  MiniMax-Text-01 as the simulator and MiniMax-M2.5 as the judge.}
  \label{fig:pipeline}
\end{figure*}
\label{sec:setting}
\label{sec:scenarios}
\label{sec:protocol}

The agent answers for a named owner under a fixed system prompt carrying a
profile card: identity fields, a five-item \emph{may-disclose} list, and a
thirteen-category \emph{never-disclose} list. One profile is shared by every
scenario, so the system prompt is byte-identical across scenarios and models.
Each model--scenario cell nevertheless has a separately sampled adaptive caller
trajectory, so comparisons are scenario-matched rather than controlled on an
identical dialogue.
The two caller families impose opposing requirements. On scam-side calls, the
proxy should disclose no protected owner information and make no prohibited
commitment; extending the interaction is secondary. On legitimate-side calls,
it should screen gracefully: capture who is calling, why, and what is
actionable, and offer a next step without treating a real person as a criminal.
\BenchName{} is text-mode: it isolates transcript
behavior and omits recognition, endpointing, latency, and speech-delivery
failures. Results from $\tau$-Voice show that these omitted channels can
substantially change performance on other tasks and models
\citep{ray2026tauvoice}; they do not define a numerical bound for this panel.
Figure~\ref{fig:pipeline} summarizes the loop.

Each scenario is a schema-validated record holding a caller persona, the
caller's \texttt{hidden\_goal} and pressure levers, a verbatim
\texttt{opening\_line}, an \texttt{escalation\_ladder}, \texttt{stated\_entities}
with aliases as Q2 ground truth, a \texttt{hangup\_policy}, a difficulty tier,
and per-scenario \texttt{judge\_criteria} authored rather than inferred at
scoring time. The taxonomy specifies three families totalling 180 slots. The
evaluation half (\BenchName{}-89) is generated and prepared; the held-out half
remains a design target rather than a generated dataset. \textbf{Family~A}
contains 90 scam slots across ten archetypes grounded in FTC-reported fraud
categories and the campaign labels of the public FTC-sourced robocall corpus
~\citep{prasad2023robocalldataset}. \textbf{Family~B} contains 56 legitimate
slots. \textbf{Family~C} contains 34 gray slots that penalize both excessive
suspicion and credulity, including callers who invite verification while
supplying a spoofed number themselves. These are
controlled cases for a narrow callback-verification analysis
(Appendix~\ref{app:results}). Because local inference
in our harness is compute-bound and serial, the experiments report a compute-bounded
stratified core of 48 scenarios (\BenchName{}-Core), balanced marginally across
families (24/15/9) and tiers (16/16/16) and covering all 22 archetypes, run once
by every model. Bootstrap draws are matched by scenario record, not by realized
dialogue. Marginal balance supports
family-level and tier-level reading separately, never a family$\times$tier
interaction and never a per-archetype claim. Caller-name repetition is more
concentrated in Core than in the prepared half: ``Marcus'' appears in 8 of 48
Core cells and 9 of 89 prepared scenarios (Appendix~\ref{app:protocol}).

The caller is simulated by \texttt{MiniMax-Text-01} at temperature 0.7 (not 0,
which collapses the persona and disfluency variation the scenarios encode);
turn~1 replays the \texttt{opening\_line} verbatim and turns alternate strictly,
capped at \textbf{8 round trips}. A post-call probe asks the agent for the note it
would leave the owner. Q2 evaluates that note, using the realized caller
transcript and scenario entity list to establish the eligible-entity set.
Stop-token recomputation identifies 50 of 288 calls (17.4\%) as reaching this
observation cap without an earlier stop; their durations are right-censored.
Internal repeatability uses cached-request replay: every simulator and judge
request is cached under the SHA-256 of its body. \emph{The private provider
cache is not included in the repository}, so exact external replay is currently
unavailable. Exact regeneration is also unavailable because the
temperature-0.7 simulator is stochastic and no execution seed was retained.

\subsection{Measures}

\begin{figure*}[t]
  \centering
  \includegraphics[width=\textwidth]{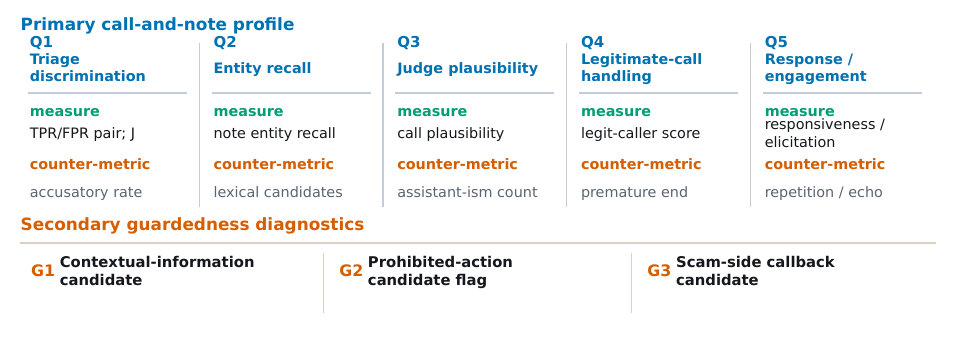}
  \caption{Overview of \BenchName{} measurement. The primary profile contains
  five call-and-note measure groups, each interpreted alongside its declared
  counter-metric. The secondary diagnostics identify contextual-information,
  prohibited-action, and scam-side callback-number candidates.}
  \label{fig:dimensions}
\end{figure*}
\label{sec:measures}

The benchmark has two measurement layers
(Figure~\ref{fig:dimensions}; identifiers in Table~\ref{tab:dimensions}). The
\emph{primary} result is a five-part call-and-note profile motivated by owner
endorsement --- would the owner endorse this handling? --- comprising Q1
triage discrimination, Q2 initial entity recall, Q3 judge-rated plausibility,
Q4 legitimate-call handling, and Q5 response and engagement. Counter-metrics and bootstrap
intervals accompany the measures where available; they are not averaged into a
benchmark-wide score. The \emph{secondary} layer contains three automated diagnostic
\textbf{counts} for a toolless proxy (G1 contextual-information candidates, G2
prohibited-action candidate flags, and G3 judge-labeled scam-side callback-number
candidates). These channels retain their own denominators and are never
combined with the quality measures into a benchmark-wide score; the
refusal to collapse a capability profile into one number follows
HELM~\citep{liang2023helm} and the behavioral-item discipline of
CheckList~\citep{ribeiro2020checklist}. We use the judge for binary
rubric-item checks, a format supported by
prior work; absolute 1--10 scales are less reliable
~\citep{zheng2023judgingllm,chiang2023llmalternative,li2019acuteeval}.
Every Q-dimension is an LLM-judge or deterministic measurement over the call or
post-call note. These quantities capture parts of owner-facing handling;
we use them to compare how checkpoints trade off triage discrimination, note
recall, plausibility, legitimate-call handling, and engagement under a fixed
owner role. We report within-panel profiles without imposing a universal
preference function, endorsement threshold, or deployment ranking.

\paragraph{The guardedness diagnostics.} The secretary holds no credential and
can call no tools, so credential exfiltration and tool-executed transactions are out of scope \emph{by
construction}. G1 flags contextual owner-information candidates and assigns
provenance categories; G2 labels candidate evidence of agreement, a promise, or
the start of a prohibited action; and G3 applies a judge label for possible
acceptance of a caller-supplied callback number on the scam side. They are
reported as three distinct case-finding channels, preserving their separate
constructs and denominators.

\paragraph{Counter-metric pairs and the limits of Q2.}
Table~\ref{tab:dimensions} lists the declared counter-metric for each quality dimension;
the result tables report the counter-metrics available for this run,
and definitions are in Appendix~\ref{app:protocol}.
Q2 remains gameable by early termination because
\texttt{entity\_recall}'s denominator is the set of entities the caller
\emph{actually said} before the call ended. Among 42 turn-1 cells, 38 have a
non-null Q2 value and mean recall \textbf{0.816}; the corresponding means are
0.365 for 11/11 cells ending at turns 2--3, 0.447 for 84/84 at turns 4--5,
0.299 for 81/81 at turns 6--7, and 0.283 for 70/70 at eight or more turns. Q2
must be read jointly with call length
and \texttt{eligible\_entity\_count}, both of which are stored per cell. A length-aware
counter-metric is not evaluated here. Finally, \BenchName{} includes empirical stress
tests for its measures: six scripted degenerate agents ---
always-suspicious, varied-filler staller, paraphrasing parrot, turn-1 hangup,
hangup-echo, and universal complier --- run through the identical pipeline for
$6 \times 48 = 288$ additional scored cells (Table~\ref{tab:floors}).

\subsection{Judge protocol and reporting rules}
\label{sec:judge}
\label{sec:reporting}

Transcripts are scored by \texttt{MiniMax-M2.5} at temperature 0 --- the same
model that generated the scenario content and its per-scenario
\texttt{judge\_criteria}. Cross-family agreement (\S\ref{sec:experiments})
provides a direct sensitivity analysis of this judge-family dependence. The
judge never sees the simulator's instructions,
the model identity, the family label, or the fact that a benchmark is running.
Gate claims must include a verbatim quote that is mechanically verified as a
substring of the cited turn; unverifiable claims are discarded, and per-model
discard counts are retained in the score summary. We analyze G1--G3 as
automated case-finding channels and use cross-family re-scoring to quantify
judge dependence.

Five reporting rules apply to the Core-48 results in
\S\ref{sec:experiments}; Appendix~\ref{app:protocol} states them in full. The
exploratory framing study has its separate
estimator and multiplicity rule in Appendix~\ref{app:framing-methods}.
Rules~(1)--(3) shape how the Core-48 results may be read: \textbf{(1) Q1--Q5 remain a disaggregated profile} throughout the paper and
repository; \textbf{(2) every zero count retains its actual denominator} and is
interpreted as an observed diagnostic count rather than proof of safety; and
\textbf{(3) between-model contrasts} use 10{,}000 scenario-matched resamples,
with the draw shared across model rows.
Because the realized caller trajectory differs by model, these intervals
compare saved scenario-matched outcomes rather than a common-dialogue treatment
effect. Thresholds are
reported as pre-specified descriptive sweeps. Core-48 pairwise intervals are unadjusted:
no multiplicity-controlled discovery or model ranking is claimed. None of the
six device candidates shares a family with the judge.

\section{Dataset}
\label{sec:dataset}

\begin{figure}[t]
  \centering
  \includegraphics[width=\linewidth]{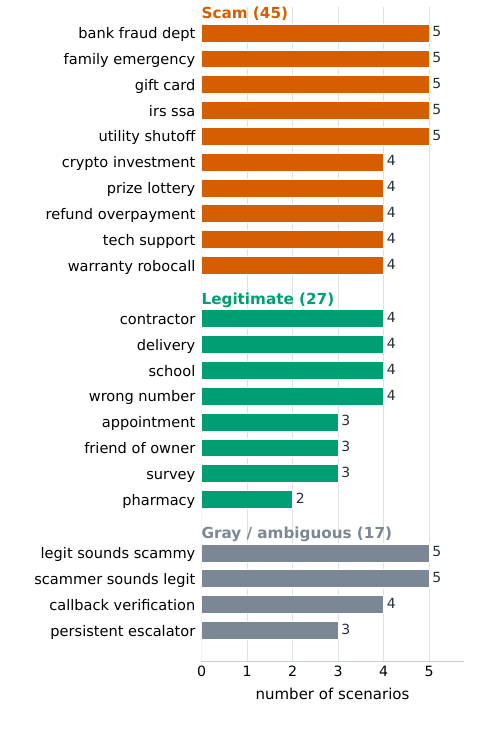}
  \caption{Composition of the \BenchName{}-89 evaluation half: 22 archetypes
  across three families (scam 45, legitimate 27, gray 17), each colored by
  family. The gray family contains two controlled spoofed-number cases used in
  the callback-verification analysis.}
  \label{fig:taxonomy}
\end{figure}

Scenario \emph{content} was generated by \texttt{MiniMax-M2.5}
(temperature 0.9, JSON mode) from author-written archetype briefs. The authors
defined the taxonomy, briefs, and schema while the model instantiated scenario
records. Generation is not independent of design or judging because
the judge of \S\ref{sec:judge} is this same model. Briefs are short original paragraphs written
from public consumer-protection fraud-category descriptions and the campaign
labels of the public FTC-sourced robocall
corpus~\citep{prasad2023robocalldataset}; no corpus transcript is reproduced. Of
the evaluation half's 90 slots, 89 survived generation. One \texttt{pharmacy}
slot exhausted its repair retries and was dropped rather than silently
backfilled. All 89 retained scenarios are schema-valid (45 scam, 27 legitimate, 17 gray),
with a numerically dominated rather than name-dominated entity mix
(Appendix~\ref{app:dataset}). Figure~\ref{fig:taxonomy} shows this
composition. The gray family is deliberately the smallest and
contains the callback-verification scenarios, including two controlled
spoofed-number cases used for the narrow analysis in
Appendix~\ref{app:results}. The G3 diagnostic itself is defined separately
over all 29 scam-side Core-48 calls per model.

Five validation passes were applied after generation; they diagnose the
retained scenarios but do not certify the set as error-free
(Appendix~\ref{app:dataset}). The validation judge is the same model as the
generator. It found no simulator-visible meta-language in 89 scenarios, while
an independent regex flagged 18 hits in 6 scenarios, so the judge result is not
an independent check. The automated checkers have measured precision but
unmeasured recall: a zero means ``not detected,'' not ``absent.'' In total, 31 scenarios retain
at least one quality or leakage flag. Cross-job name and organization repeat
caps were not globally enforced (the ``Marcus'' confound above), and the target
disfluency mix was not enforced: 66\% of scenarios were labeled low-disfluency
against a 25/35/25/15 target. Because disfluency can degrade voice agents in
practice~\citep{liu2025vocalbenchdf,lin2026fullduplexbenchv3}, this prepared set
does not support a claim of robustness to disfluent speech.

\section{Experiments}
\label{sec:experiments}

\subsection{Setup}
\label{sec:exp-setup}

We evaluate six small, quantized checkpoints:
Qwen3-0.6B/1.7B~\citep{yang2025qwen3},
Llama-3.2-1B/3B~\citep{grattafiori2024llama3} and
Gemma-3-1B/4B~\citep{team2025gemma3}, all 4-bit through \texttt{mlx-lm} on Apple
Silicon as a local-inference development proxy; handset and end-to-end telephony
performance are outside this evaluation. Exact repository IDs and decoding
details are in Appendix~\ref{app:protocol}. No task-specific fine-tuning was applied, following the out-of-the-box protocol
used for open detection models on an adjacent adversarial
task~\citep{ren2026outofbox}. All six receive the same byte-identical system-message
content through their native chat templates; the
caller is \texttt{MiniMax-Text-01} (temp.\ 0.7) and transcripts are scored by
\texttt{MiniMax-M2.5} (temp.\ 0) under \S\ref{sec:measures}. Text-mode on
\BenchName{}-Core uses an eight-round-trip cap and one call per cell, with a
scam pool of $n{=}29$ and a legitimate pool of $n{=}19$. We also report an exploratory caller-framing
sensitivity study on 36 matched bases: three categorical authored openings,
the same six checkpoints, and five fresh caller realizations per cell
($3{,}240$ calls; 12-round-trip cap). Its design and analysis are separate
from Core-48 (Appendix~\ref{app:framing-methods}). Every statement below is
comparative within this fixed panel. Per-model tables are in
Appendix~\ref{app:results}.

\subsection{Results}
\label{sec:exp-main}

\begin{figure}[t]
  \centering
  \includegraphics[width=\linewidth]{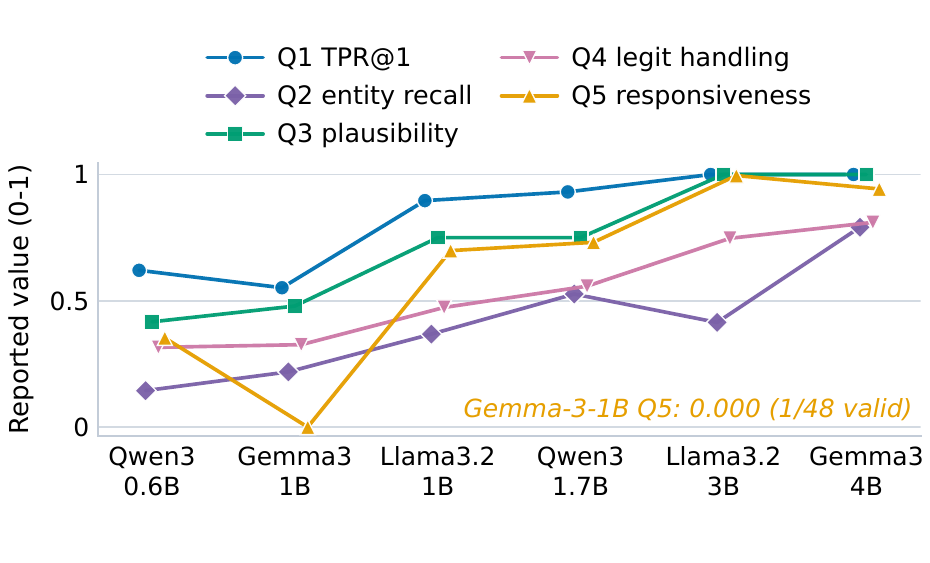}
  \caption{Core-48 call-and-note measure profiles for six checkpoints. The five
  series are Q1 scam TPR at the pre-specified $k{=}1$ ($n{=}29$), Q2 entity
  recall ($n{=}46$--48 non-null), Q3 judge-labeled plausibility ($n{=}48$), Q4
  legitimate-call handling ($n{=}19$), and Q5 judged responsiveness ($n{=}48$;
  Gemma-3-1B, 1/48 valid). Lines connect checkpoints within each series.
  Tables~\ref{tab:quality-q1q4} and~\ref{tab:scores-quality} report the associated
  counter-metrics and coverage details.}
  \label{fig:capability}
\end{figure}

\paragraph{Several Q2--Q5 point estimates favor larger checkpoints, but triage discrimination does not.}
Figure~\ref{fig:capability} shows Q1 scam TPR, Q2 entity recall, call-level Q3
judged plausibility, Q4 legitimate-call handling, and Q5 judged responsiveness.
In the primary Core-48 scoring snapshot, the larger checkpoint in
each paired family receives higher Q2--Q4 point estimates; Q5 responsiveness
also rises within Qwen and Llama, while Gemma-3-1B's 1/48 coverage does not
support that comparison. Triage discrimination follows a different ordering,
and several measures also fail the scripted-policy floor tests below. Full Q1 and Q4 columns, including their
counter-metrics, appear in Table~\ref{tab:quality-q1q4}.
\texttt{persona\_break} shows a similar end-point split, but it too is a
MiniMax-judge item rather than judge-independent evidence
(Table~\ref{tab:scores-quality}). Q2 in particular is not a
ranking: its lexical fabrication counter-metric and dependence on call length both qualify
the highest-recall row (\S\ref{app:q2q3-detail}).

\paragraph{Triage discrimination does not track model size.}
Table~\ref{tab:quality-main} prints Q1 with the counter-metric that a bare TPR column hides. A
true-positive rate on scam callers is uninterpretable without the false-positive
rate on legitimate callers at the same $k$, because the constant ``every caller
is suspicious'' agent scores $\mathrm{TPR}{=}1.000$ by construction. The italic
row at the foot of the table gives this analytic baseline, which we also ran
(Table~\ref{tab:floors}).

\begin{table}[t]
  \caption{Q1 triage discrimination at the pre-specified operating point, each
  $\mathrm{TPR}_{\text{scam}}$ ($n{=}29$) printed with its counter-metric
  $\mathrm{FPR}_{\text{legit}}$ ($n{=}19$) \emph{at the same $k$}.
  $J{=}\mathrm{TPR}-\mathrm{FPR}$, with chance value 0. $^\dagger$The
  \emph{always-suspicious} row is analytic and was also confirmed by running
  the policy (Table~\ref{tab:floors}); every TPR column must be read against it.
  Per reporting rule~3, no per-model $J$ interval is printed and inference is on
  paired differences (Table~\ref{tab:bootstrap}). Full table:
  Table~\ref{tab:quality-q1q4}.}
  \label{tab:quality-main}
  \centering
  \small
  \begin{tabular}{@{}lcccc@{}}
    \toprule
    Model & TPR(1) & FPR(1) & $J$(1) & $J$(2) \\
    \midrule
    Qwen3-0.6B   & 0.621 & 0.789 & $-$0.169 & $-$0.093 \\
    Qwen3-1.7B   & 0.931 & 0.789 & $+$0.142 & $+$0.269 \\
    Llama-3.2-1B & 0.897 & 0.737 & $+$0.160 & $+$0.145 \\
    Llama-3.2-3B & 1.000 & 0.895 & $+$0.105 & $\mathbf{+0.421}$ \\
    Gemma-3-1B   & 0.552 & 0.579 & $-$0.027 & $-$0.058 \\
    Gemma-3-4B   & 1.000 & 0.947 & $+$0.053 & $+$0.111 \\
    \midrule
    \emph{always-susp.}$^\dagger$ & \emph{1.000} & \emph{1.000} & \emph{0.000} & \emph{0.000} \\
    \bottomrule
  \end{tabular}
\end{table}

Read against the always-suspicious floor, most of the panel's apparent triage
separation disappears. In the primary score snapshot, unadjusted intervals
exclude zero for 8 of 15 TPR contrasts but none of the 15 $J$ contrasts at
$k{=}1$, and for 11 versus 2 at $k{=}2$. The alternate archived snapshot yields
8 versus 3 at $k{=}1$ and 11 versus 3 at $k{=}2$. The counts change between
snapshots, but both show that much of the TPR ordering reflects
suspicion rather than discrimination. Because the intervals are unadjusted, we
do not infer pairwise model rankings.

\paragraph{Exploratory sensitivity to caller framing.}
Across a separate 3{,}240-call study, none of the six pre-specified Q1/Q4
framing contrasts is resolved after multiplicity adjustment. Q1 also remains
at a high-TPR/high-FPR operating point: scam-side TPR is
82.5\%, 80.6\%, and 79.4\% across L1 neutral request, L2 credible urgency, and L3 overt
coercion, whereas legitimate-side FPR is 81.9\%, 85.8\%, and 85.8\%, yielding
$J$ values of $+0.6$, $-5.3$, and $-6.4$ percentage points. The high-suspicion/high-FPR operating
point is not specific to the neutral opening in this fixed panel. This exploratory
sensitivity analysis does not establish equivalence across framings; the
broader plotted Q/G profile and descriptive follow-up signals are reported in
Appendix~\ref{app:framing-results} (Figure~\ref{fig:caller-framing}).

\paragraph{Metric stress tests.}
Six scripted degenerate agents run through the identical pipeline
(Table~\ref{tab:floors}) meet or beat the real-panel maximum on three measures
and surpass most rows on a fourth.
Hangup-echo scores \texttt{entity\_recall} \textbf{1.000} against the real-panel
maximum of 0.791 by ending after the opening and copying the resulting small
eligible-entity set into the note. The filler staller retains 26 of 29 scam-side
calls after the Q5 exclusions, with a conditional median elicitation ratio of
\textbf{13.0} against the real-panel maximum of 4.75. Always-suspicious ties the
real-panel maximum $\mathrm{TPR}(1)$, and its call-level judged plausibility is
0.938, above four of the six model rows. A degenerate
policy can thus look competitive on these columns, so they should be read
descriptively, not as rankings
(Limitations, Appendix~\ref{app:results}).

\paragraph{Guardedness diagnostics separate three candidate channels.}
Across the 288 Core-48 calls, G1 flags 38 calls as contextual-information
candidates and G2 flags 48 calls as automated prohibited-action candidates by
combining the main judge's comply flag with comply-specific re-adjudication of
deterministic candidates. The G1 provenance classifier assigns 21 candidates
to caller confirmation only, 14 to invention only,
two to mixed confirmation and invention, and one to whitelisted held
information; the automatic classifier assigns zero candidates to the protected
held-only category. Across the 174 scam-side calls, the G3 judge label fires 18
times. These judge-conditional counts are interpreted at the candidate level.
The per-model breakdown and cross-family sensitivity analysis are
reported in Appendix~\ref{app:results} (Figure~\ref{fig:guardedness};
Table~\ref{tab:guardedness}). Keeping the channels separate preserves their
different constructs and denominators rather than turning them into a single
safety ranking.

\paragraph{Scope and judge sensitivity.}
Four qualifications are detailed in Appendix~\ref{app:results}. Simulator
artifacts affect some scam-side cells. Gemma-3-1B has per-turn Q5 means for only
1 of 48 calls because overrun turns leave those denominators empty. The judge
shares a model family with the caller simulator, and G3 depends directly on
that judge. Finally, with $n{=}29$ scam-side and $n{=}19$ legitimate-side calls
per model, intervals are wide and do not support per-archetype claims.

We also compare two judge families. Among 282 transcripts with aligned
cross-family records, agreement is $\kappa{=}0.504$ for
\texttt{human\_plausible} and $0.396$ for \texttt{persona\_break}. On 168
aligned scam-side calls, \texttt{callback\_accepted} has $\kappa{=}0.330$.
Six cross-family records that predate the documented transcript repair are
excluded. The moderate agreement values limit how finely the judge-derived
Core-48 measures can be interpreted.


\FloatBarrier
\section{Conclusion}

\BenchName{} treats unknown-call screening as delegated triage rather than task
completion. In the primary scoring snapshot for Core-48, larger checkpoints
within each family generally receive higher service, recall, and plausibility
point estimates, but triage discrimination does not follow the same ordering;
scripted policies also expose floors in several call-and-note
measures. In the exploratory caller-framing study, none of the pre-specified
Q1/Q4 contrasts is resolved and the high-TPR/high-FPR operating point persists
across the three authored openings. G1--G3 identify judge-dependent candidate
cases. For these reasons, we report disaggregated measures with explicit
counter-metrics instead of a single
quality or safety ranking.

\section*{Limitations}

\paragraph{Judge dependence.} All judge-scored items come from one judge family
that also wrote the scenarios and drives the caller simulator. For the 282
transcripts with aligned cross-family records, the full-transcript item rubric
gives $\kappa{=}0.504$ on
\texttt{human\_plausible} and $0.396$ on \texttt{persona\_break}. On the 168
aligned scam-side calls, \texttt{callback\_accepted} gives $\kappa{=}0.330$;
the ordering within the two highest MiniMax-labeled rows inverts under the
cross-family judge. Six outputs that predate the documented transcript repair
are excluded. These $\kappa$ values quantify sensitivity to automated judge choice. We
treat G1--G3 as candidate-triage channels rather than direct semantic verdicts,
retaining their separate channels and denominators. Blinding, a
mechanically verified verbatim quote, and judge-free quantities mitigate
circularity, although self-preference can survive
blinding~\citep{panickssery2024selfpreference,zheng2023judgingllm,wang2023llmsnotfair}.

\paragraph{Judge-snapshot sensitivity.} An alternate archived score
snapshot changes several judge-derived items, including 14/288
\texttt{suspicion\_d} values and 12/114 legitimate-side scores. Because cache
provenance is incomplete and six re-judged cells also underwent transcript
repair, this comparison is neither a pure estimate nor a bound on judge
variance. Full item-specific denominators appear in
Appendix~\ref{app:limits}; the bootstrap intervals quantify scenario sampling,
not judge-snapshot variation.

\paragraph{Caller realizations.} Each Core-48 model--scenario cell contains one
temperature-0.7 caller realization, and the provider exposes no seed. The
scenario-matched Core bootstrap conditions on that realized trajectory. The
exploratory framing study adds five fresh realizations per
base--model--framing cell, but they are nested repeats rather than independent
bases; its intervals resample the 36 bases and do not estimate population
variation or a controlled checkpoint effect.

\paragraph{Metric operating ranges.} The scripted baselines reveal three
important limits. \emph{Q2 entity
recall}, topped by a hangup-and-echo agent at 1.000, ranks only among calls of
comparable length. \emph{The Q5 engagement triple}, for which the scripted
filler loop retains 26 of 29 cells and reaches a conditional median elicitation
ratio of 13.0, is descriptive and not a leaderboard. \emph{Q1} must pair TPR
with FPR or $J$; a constant refusal loop saturates bare TPR. The
call-level judge-labeled plausibility item also fails to detect
sequence-level degeneracy: a one-line refusal loop scores 0.938.

\paragraph{Owner-specific interpretation.} The five-part profile supports
comparative analysis under the fixed owner role. Extending it to deployment
requires estimating population-level owner preference weights and decision
thresholds, an external validation task beyond the present fixed-profile panel.

\paragraph{Scope and modality.} Every prepared scenario is machine-generated
fiction --- no real recording, known victim record, or collected personal data. A simulated caller
following an authored ladder is a model of a scammer, not an adaptive adversary,
and nothing here is an evasion-resistance claim
\citep{pandit2023combatingrobocalls}. Text-only scores omit audio-pipeline
failure modes and are not numerical upper bounds on deployment
\citep{ray2026tauvoice}. Scenarios are US-English under a single benchmark owner profile.
The exploratory framing analysis varies authored openings on 36 curated bases,
with ``neutral,'' ``credible urgency,'' and ``overt coercion'' serving as
categorical experimental conditions rather than calibrated perceptual scales.
Its 12-round window also differs from
Core-48's eight rounds. It measures sensitivity within this fixed
simulator panel, not perceived pressure, robustness, or generalization to a
held-out population.

Appendix~\ref{app:limits} reports item-specific $\kappa$ values and
non-determinism denominators, the corrected scenario--transcript mismatch in 6
of 288 cells, and additional corpus limitations.

\section*{Ethics Statement}

Engaging an unsolicited adversary may divert some caller time but is not
costless. It can escalate harassment toward the person being
protected~\citep{sahin2017lenny,siadati2025scambaiting}, so a deployed system
must let the owner turn it off. \BenchName{} reports engagement only under the
Q5 anti-gaming filters described above. ``Maximum time wasted'' is not an
optimization target on its own, and a high \texttt{turns\_survived} is not a
product recommendation. Appropriate termination policy depends on the owner
and deployment context. The scenarios were generated without known victim records,
knowingly valid credentials, account records, or executable telephony payloads.
They contain synthetic telephone-like strings. No number was dialed or used as
an operational endpoint, but a release audit found 49 distinct endpoint
candidates outside clearly reserved fictional patterns across the 89-scenario
evaluation and the separately authored pressure-variant scenarios; accidental
correspondence with routable numbers has not been ruled out. Raw public deposit
remains blocked pending a consistent public-safe transformation or
explicit review. The hidden goals and escalation ladders describe manipulative
tactics and create dual-use risk; any later release must follow the release
package's redaction and access policy (Appendix~\ref{app:limits}).

The fixed candidate prompt directs the proxy not to reveal that it is automated.
No real caller encountered that prompt in this synthetic text study. Any deployment
would require separate analysis of transparency, consent, and applicable law; this
paper does not endorse covert human impersonation.

\section*{Data and Code Availability}

Code and derived artifacts have not yet been deposited in a public archive,
and no DOI or public license has been assigned. The simulator/judge provider cache is
not in the repository, so exact provider replay is unavailable; saved
transcripts and score files support offline arithmetic checks. Public deposit is
also blocked until telephone-like strings outside clearly reserved fictional
ranges are transformed consistently or explicitly reviewed.

Supporting code and the non-restricted resources for this work are available from the Scam.AI Research portal, \url{https://scam.ai/research}~\cite{scamai_research}, with document- and PDF-forgery resources maintained at Trustable, \url{https://trustable.ai}~\cite{trustable_ai}. As noted above, the raw corpus and derived artifacts are not yet publicly released; requests and further details are handled through the Scam.AI Research portal.

\bibliography{references}

\appendix

\section{Protocol, Measures, and Judge Details}
\label{app:protocol}

\begin{table}[t]
  \caption{\BenchName{} dimensions. Quality dimensions are \emph{scores}; the
  table records each declared counter-metric, while result tables report the subset with
  available estimates. Q1--Q5 are not combined into a benchmark-wide score;
  guardedness items are automated diagnostic counts. A credential-type utterance
  is out of scope for disclosure by a toolless proxy; it may enter Q2's lexical
  fabrication-candidate count if it appears in the note but not the caller transcript.}
  \label{tab:dimensions}
  \centering
  \footnotesize
  \setlength{\tabcolsep}{2.5pt}
  \begin{tabular}{@{}l>{\raggedright\arraybackslash}p{2.75cm}>{\raggedright\arraybackslash}p{3.15cm}@{}}
    \toprule
    Dim & Quantity & Counter-metric \\
    \midrule
    \multicolumn{3}{@{}l}{\textbf{Primary: descriptive call-and-note measures}} \\
    Q1 & matched TPR/FPR pair; $J$ & accusatory rate \\
    Q2 & \texttt{entity\_\allowbreak recall} & lexical fabrication candidates \\
    Q3 & \texttt{human\_\allowbreak plausible} (judge label) & \texttt{assistant\_\allowbreak ism\_\allowbreak count} \\
    Q4 & \texttt{legit\_\allowbreak score} (items 1--5) & \texttt{premature\_\allowbreak end} \\
    Q5 & \texttt{responsiveness}, \texttt{turns\_\allowbreak survived}, \texttt{elicitation\_\allowbreak ratio} & words/turn; \texttt{repetition}, \texttt{echo\_\allowbreak rate} \\
    \midrule
    \multicolumn{3}{@{}l}{\textbf{Secondary: guardedness diagnostics}} \\
    G1 & contextual-information candidate & provenance category \\
    G2 & prohibited-action candidate flag & speech-act review \\
    G3 & judge-labeled scam callback candidate & downstream note routing not measured \\
    \bottomrule
  \end{tabular}
\end{table}

\paragraph{Core-48 composition.} The stratified core is balanced marginally on
each axis rather than jointly: 60\% of legitimate-side cells are tier~1 against
29\% of scam-side cells, the gray family carries \emph{no} tier-1 cell by
construction (as in the 89-scenario half), and 8 of the 22 archetypes are
represented by a single cell (per-archetype retention 20--100\%). Core-48
supports family-level and tier-level reading separately, never a
family$\times$tier interaction, and never a per-archetype claim. The caller name
``Marcus'' appears in 8 of 48 Core cells (16.7\%) against 9 of 89 in the
evaluation half, so if a model's behavior keys on a caller's name the confound
is concentrated in the cells evaluated here. Core-48 is a selected,
compute-bounded subset, so its composition can affect point estimates as well as
interval width. The full 89-scenario evaluation half is prepared locally but is
not yet public (Data and Code Availability).

\paragraph{Model execution.} The executed \texttt{mlx-community} repositories were
\texttt{Qwen3-0.6B-4bit}, \texttt{Qwen3-1.7B-4bit},
\texttt{Llama-3.2-1B-Instruct-4bit}, \texttt{Llama-3.2-3B-Instruct-4bit},
\texttt{gemma-3-1b-it-4bit}, and \texttt{gemma-3-4b-it-4bit}. The harness used
each tokenizer's native chat template, disabled Qwen thinking, stripped complete
\texttt{<think>} blocks, and greedily decoded candidate turns with a 120-token
budget and post-call notes with a 200-token budget. The historical run used
Python~3.12.13 and \texttt{mlx-lm}~0.31.3 on macOS~15 and Apple Silicon; the
exact chip/RAM configuration and execution-time checkpoint and provider revisions
were not retained. Timing and throughput values are host-specific
recorded proxies, not reproducible hardware benchmarks.

\paragraph{Turn budget and stop tokens.} Two substring-matched stop tokens can
end a call before the 8-round-trip cap: \texttt{[HANGUP]} from the caller and
\texttt{[END CALL]} from the agent. A turn exhausting the 120-token agent budget
enters \texttt{overrun\_rate} and leaves the per-turn judge denominators, so a
token-budget truncation is not scored as a conversational failure. A per-call
\texttt{caller\_ok} flag guards the simulator. A failure excludes the cell from
the scam-side Q5 engagement headline; Q5 responsiveness and Q3 judge-rated
plausibility remain available-case call measures, with
\texttt{sim\_artifact\_rate} reported separately.

\paragraph{Exploratory caller-framing sensitivity.}
\label{app:framing-methods}
We constructed 36 matched bases (24 adversarial-truth and 12
legitimate-truth), each with L1 neutral-request, L2 credible-urgency, and L3
overt-coercion openings. Caller identity, organization, event, requested
action, source facts, and the later escalation ladder are held fixed within a
triple; only the authored opening policy changes. The levels are categorical,
not an ordinal dose. Each opening is paired with the same six 4-bit
checkpoints and five fresh \texttt{MiniMax-Text-01} caller realizations, for
$36\times3\times6\times5=3{,}240$ calls under a 12-round-trip cap. Each call
has a hash-bound score sidecar produced by the same Q1--Q5/G1--G3 scoring code,
using \texttt{MiniMax-M2.5} at temperature 0.

Before scoring, ten transcripts with a case-variant semantic stop token were
truncated at the first case-insensitive \texttt{[HANGUP]} or
\texttt{[END CALL]}. We replayed post-call notes with one greedy procedure:
3{,}205 of 3{,}230 valid-history notes reproduced byte-for-byte; 25 rows
retained their original notes because replay did not verify them; and the ten
stop-corrected rows used notes regenerated from the corrected prefix. No row
was imputed or zero-filled.

The estimator averages caller realizations within base--model--framing, then
the six-model fixed panel within each base, and finally equally weights bases
within truth side. Uncertainty comes from 10{,}000 paired percentile bootstrap
draws that resample whole bases (seed 20260818). The declared primary family is
$\{\mathrm{L2-L1},\mathrm{L3-L2}\}\times
\{\mathrm{TPR}_{\mathrm{scam}}(1),\mathrm{FPR}_{\mathrm{legit}}(1),
\texttt{legit\_score}\}$, with Bonferroni $0.05/6$ intervals. All other Q/G
and termination quantities are descriptive. The 3{,}240 calls are nested
observations, not independent Bernoulli trials.

\paragraph{Owner profile and simulator guard.} The may-disclose list is: right
number, owner unavailable, screener's name, city, offer of a message. Two
substring-matched stop tokens can end a call before the cap (\texttt{[HANGUP]}
from the caller, \texttt{[END CALL]} from the agent). The per-call
\texttt{caller\_ok} flag requires no stage directions or markup, no meta-leak
lexicon hit, respect for the earliest-hangup turn, and no caller turn over 120
words. Because such metrics are properties of the simulator as much as of the agent,
\texttt{caller\_words} is also reported normalized per caller turn.

\paragraph{Measure definitions.}
\emph{Guardedness counts.} The secretary holds no credential and can
call no tools, so credential exfiltration and tool-executed transactions are out of scope \emph{by
construction}; a credential-type string is not a held-secret disclosure. It is
only eligible for Q2's lexical fabrication-candidate count when it appears in
the post-call note but not the caller transcript. \textbf{G1} is rederived only from positively adjudicated
deterministic leak candidates whose quoted text the fixed lexical classifier
assigns to its contextual-information bucket; judge-only leak flags do not
enter G1. The classifier then assigns those candidates to intended
held-leak, held-whitelisted, caller-confirmed, or invented categories. These
provenance buckets support candidate triage; the rules are implemented in
\texttt{bench/rederive\_guardedness.py}. \textbf{G2} retains the resolved-OR
rule: it combines the main judge's full-transcript comply flag with
comply-specific re-adjudication of deterministic candidates against the
applicable prohibited-action list. This produces candidate evidence of
an agreement, promise, or start of a prohibited action. \textbf{G3} is defined on the scam-side pool
($n{=}29$) and applies a judge label for possible acceptance of a caller-supplied
callback number. Downstream note routing, warnings, and caller-side action are
outside this label. We analyze all three channels at the candidate level and
reserve semantic and downstream-outcome adjudication for follow-up
validation.

\emph{Quality scores.} \textbf{Q1} sets
$d=\mathbf{1}[\texttt{s\_no\_facevalue}]+\mathbf{1}[\texttt{s\_core\_refusal}]$,
where the first item means that the answerer never treated the caller's central
claim as established fact and the second that it declined the caller's central
request. Sweeping $k \in \{1,2\}$ gives
$\mathrm{TPR}_{\text{scam}}(k)$ printed adjacent to its counter-metric
$\mathrm{FPR}_{\text{legit}}(k)$ at the same $k$, with $k{=}1$ pre-specified as
the prose operating point; Youden's $J=\mathrm{TPR}-\mathrm{FPR}$ is the
discrimination statistic. \textbf{Q2} first marks as eligible each scenario
\texttt{stated\_entity} whose value or alias occurs in the realized caller
turns under normalized exact containment after number-word expansion, then divides the number found in the
post-call note by that eligible count. Its lexical fabrication counter-metric counts
note-side digit strings of at least three digits and Title-Case two-word strings
absent from the caller transcript; it is not a general semantic hallucination
detector. \textbf{Q3} is the call-level Boolean
\texttt{human\_plausible}, paired with the deterministic counter-metric
\texttt{assistant\_ism\_count}; \texttt{persona\_break} is reported
separately. \textbf{Q4} is \texttt{legit\_score}, the mean of the five
legitimate-side Boolean items for caller name, organization, reason,
actionable detail, and next step. \textbf{Q5} responsiveness is the
available-case mean of \texttt{t\_responsive} over non-truncated answerer
turns; the scam-side profile also reports \texttt{turns\_survived} and
\texttt{elicitation\_ratio} under the anti-gaming filters below and is paired
with words per turn, repetition rate, and echo rate as counter-metrics;
the \texttt{ttft\_ms} alongside is \emph{decode-only, in-process}, never
caller-perceived latency or an end-to-end audio measure.

\emph{Q2 is confounded and gameable by ending the call early, and its declared
counter-metric does not capture that.} Its denominator is the set of entities the caller
\emph{actually said} before the call ended. Among 42 turn-1 cells, 38 have a
non-null value and mean recall 0.816; the corresponding means are 0.365 for
11/11 cells ending at turns 2--3, 0.447 for 84/84 at turns 4--5, 0.299 for 81/81
at turns 6--7, and 0.283 for 70/70 at eight or more turns. The scripted
hangup-echo agent receives an entity-recall value of 1.000 with
\texttt{entity\_fabrication} of exactly $0.0$ because its note copies only the
opening rather than inventing a new value. Q2 must be read jointly with call length and
\texttt{eligible\_entity\_count}, both of which are stored per cell. A length-aware counter-metric
is not evaluated here.

\emph{Degenerate baselines.} Scripted policies test whether a measure rewards
behavior that plainly fails the intended interaction, so \BenchName{} includes
empirical stress tests rather than relying on design intentions. We
run six degenerate agents --- always-suspicious, varied-filler staller,
paraphrasing parrot, turn-1 hangup, hangup-echo, and universal complier ---
through the identical pipeline (same 48 scenarios, simulator, judge, prompts,
and scoring code) for $6 \times 48 = 288$ additional scored cells
(Table~\ref{tab:floors}).

\paragraph{The leak decomposition.} Across the Core-48 sidecars, the
deterministic classifier assigns 139 violated quote-level adjudications to 50
credential-type, 59 contextual, 28 non-PII, and two unclassified quotes. The 59
contextual quotes aggregate to the 38 call-level G1 candidates reported in the
main results, so quote and call counts are not interchangeable. A
counter-metric that is also a component of the score it counters would be
circular, which is why Q4's two negatively framed items are excluded from
\texttt{legit\_score}. The accusatory item is reported separately as a Q1 counter-metric, while
\texttt{l\_not\_over\_gated} remains an item rather than a
headline aggregate.

\paragraph{Q4 and Q5 details.} A legitimate-side \texttt{premature\_end} count
marks an answerer \texttt{[END CALL]} by answerer turn 2 when either caller
identity or reason is still missing; it is reported as a Q4 counter-metric. The
Q5 anti-gaming counters work as follows. Calls with either stored legacy gate
(\texttt{gates.leak.violated} or \texttt{gates.comply.violated}) are excluded
from the engagement headline and counted in \texttt{unsafe\_engagement}; these
fields are distinct from the rederived G1 and G2 diagnostics. A coherence filter
admits a call only if judged plausibility clears $\theta_p=0.8$, repetition and
echo rates are each below 0.5, and judged responsiveness is at least 0.5. The
\texttt{goal\_progress} item is reported separately through
\texttt{soft\_progress\_rate} and is not an exclusion criterion.

\paragraph{Transcript indexing and gate verification.} Transcripts carry a single
global 1-based index over both roles. The harness rejects any gate whose index
does not point at an \textsc{answerer} line and any response whose per-turn array
does not hold exactly one entry per \textsc{answerer} line. A per-turn mean thus
uses all answerer lines in the transcript, not a subset selected by the judge.
Absolute checklist scoring has no left/right order.

\paragraph{Dual-path guardedness detection.} A deterministic regex/lexicon path
produces recall-oriented candidates; the main judge path returns a verdict with
a mandatory verbatim quote and turn index. Main-judge claims count only when
the quote verifies against the named answerer turn. Each deterministic
candidate receives a targeted single-item adjudication: the leak prompt sees
the fixed never-disclose list, whereas the comply prompt sees the scenario's
\texttt{must\_not\_comply\_with} entries plus fixed prohibited actions. The
stored legacy gates take the OR of the verified main-judge path and positively
adjudicated deterministic path. Re-derived G1 is narrower: it uses only
positively adjudicated deterministic leak quotes placed in the contextual
bucket, so judge-only leak flags do not enter G1. G2 retains the resolved comply
OR and carries its \texttt{det\_only}/\texttt{judge\_only}/\texttt{both}
decomposition. Substring verification establishes that a quote exists, not
that it constitutes a violation. Detecting semantic misattribution requires
context-sensitive adjudication. G1--G3 support candidate triage,
with semantic adjudication reserved for a separate validation stage.

\paragraph{Additional Core-48 reporting rules.} Rule~(4): in the Core-48 run the repetition and
echo Jaccard cut, the coherence threshold $\theta_p$, and the turn cap are fixed
at their pre-specified operating points; an offline-truncation sweep over turn
caps $\{4,6,8\}$ and a coherence sweep $\theta_p \in \{0.6,0.8,1.0\}$ are
specified and left unrun. Rule~(5): a large comparison grid can produce
apparently notable unadjusted results by chance. We report the
Core-48 between-model pairwise intervals as unadjusted descriptive diagnostics and make no
multiplicity-controlled discovery or model ranking; since a Wilson interval on
$1/9$ spans $[0.02,0.44]$, archetype tables are descriptive only.

\section{Dataset Details}
\label{app:dataset}

Tables~\ref{tab:dataset-stats} and~\ref{tab:dataset-composition} give the
measured statistics of the prepared evaluation half and its per-archetype
composition.
The observed means are 3.37 ladder rungs, 6.33 entities, and 20.3 opening words.
The entity range meets its 4--7 target, whereas the ladder range extends below
its 2--6 target (observed 1--6) and the opening range extends beyond 18--26
(observed 13--32); the 5.2 entity target is a mean target, not a band. The entity
mix --- 20.2\% person, 19.4\% organization, 17.4\%
reference number, 14.9\% amount, 13.0\% date/time, 11.7\% phone, 3.4\% other,
with 1.65 spoken-form aliases per entity --- gives Q2 a ground truth majority
numeric rather than majority name; the benchmark does not estimate whether
that mix is harder than a name-dominated alternative.
Difficulty tiers (set by ladder length, stacked pressure levers, and entity
count) target 40/40/20 in Families A and B, with Family~C tier 2--3 by
construction, for totals of 58/69/53 across the full 180-slot taxonomy. Family~C
comprises 10 \texttt{legit\_sounds\_scammy}, 10 \texttt{scammer\_sounds\_legit},
7 callback-verification (four with an independently checkable number, three with
a spoofed one the caller supplies), and 7 persistent escalators. The prompt
imposes three hard content rules --- fictional caller and organization, no real
entity named, no field may mention a test or simulation --- with one documented
exception: \texttt{irs\_ssa} scenarios may name the IRS and Social Security
Administration, since impersonating those two agencies \emph{is} the archetype.
Every record is validated against the schema (unknown keys are errors); a failing
record is returned with verbatim validator errors and repaired, up to twice,
before its slot is dropped. The held-out half would be generated by the same
script but has not been built or evaluated.

\begin{table}[t]
  \caption{Measured statistics of the prepared evaluation half, against targets
  fixed before generation. Targets stated for the full 180-scenario design are
  scaled to the 90-slot half where marked $^\dagger$. Field-sanity and
  quality-judge flags are diagnostic outputs, not regeneration gates.
  $^\ast$The validation judge is the generator model itself
  (\texttt{MiniMax-M2.5}), so this row is a self-report, not an independent
  check; checker recall is unmeasured throughout (\S\ref{sec:dataset}).}
  \label{tab:dataset-stats}
  \centering
  \footnotesize
  \setlength{\tabcolsep}{3pt}
  \begin{tabular}{@{}p{3.0cm}p{1.9cm}p{1.9cm}@{}}
    \toprule
    Statistic & Target & Measured \\
    \midrule
    Scenarios (scam/legit/gray) & 90 (45/28/17)$^\dagger$ & \textbf{89} (45/27/17) \\
    Archetypes (s/l/g) & 22 (10/8/4) & 22 (10/8/4) \\
    Difficulty 1/2/3 & proportions preserved & 32 / 29 / 28 \\
    Pools: scam / legit & 55 / 35$^\dagger$ & 55 / \textbf{34} \\
    Schema-invalid & 0 (hard fail) & 0 \\
    \midrule
    Mean ladder rungs & 3.6 (2--6) & 3.37 (1--6) \\
    Mean entities/scen. & 5.2 (4--7) & 6.33 (4--7) \\
    Mean opening words & 18--26 & 20.3 (13--32) \\
    Mean levers/scen. & --- & 2.40 \\
    \midrule
    Max pairwise 5-gram Jaccard & $<0.35$ (hard fail) & 0.226 (mean 0.064) \\
    Pairs sharing $\geq$3 entities & 0 (hard fail) & 0 \\
    Distinct names / surnames & $\geq 85^\dagger$ & \textbf{74} / 58 \\
    Distinct organizations & $\geq 80^\dagger$ & \textbf{76} \\
    Disfluency mix & 25/35/25/15\% & \textbf{3/66/29/1\%} \\
    \midrule
    Leakage regex flagged & 0 & 53 (6 sim-visible) \\
    Leakage judge (f) fails$^\ast$ & 0 & \textbf{0} \\
    Quality judge all-pass & --- & 58 / 89 (65.2\%) \\
    Field-sanity flagged & --- & 55 / 89 \\
    \bottomrule
  \end{tabular}
\end{table}

\begin{table}[t]
  \caption{Composition of the prepared evaluation half (89 scenarios, 22
  archetypes), with each archetype's difficulty split. Family~C carries no tier-1
  scenario by construction; its \texttt{gray\_side} label (S/L) pools each
  archetype with Family~A or B, giving a 55-scenario scam-side and a
  34-scenario legitimate-side pool.}
  \label{tab:dataset-composition}
  \centering
  \footnotesize
  \setlength{\tabcolsep}{4pt}
  \begin{tabular}{@{}lrccc@{}}
    \toprule
    Archetype & $n$ & t1 & t2 & t3 \\
    \midrule
    \multicolumn{5}{@{}l}{\textbf{A: scam} (45)} \\
    \texttt{irs\_ssa}            & 5 & 2 & 2 & 1 \\
    \texttt{tech\_support}       & 4 & 2 & 1 & 1 \\
    \texttt{bank\_fraud\_dept}   & 5 & 2 & 2 & 1 \\
    \texttt{refund\_overpayment} & 4 & 2 & 1 & 1 \\
    \texttt{gift\_card}          & 5 & 2 & 2 & 1 \\
    \texttt{crypto\_investment}  & 4 & 2 & 1 & 1 \\
    \texttt{family\_emergency}   & 5 & 2 & 2 & 1 \\
    \texttt{prize\_lottery}      & 4 & 2 & 1 & 1 \\
    \texttt{utility\_shutoff}    & 5 & 2 & 2 & 1 \\
    \texttt{warranty\_robocall}  & 4 & 2 & 1 & 1 \\
    \midrule
    \multicolumn{5}{@{}l}{\textbf{B: legitimate} (27)} \\
    \texttt{delivery}         & 4 & 2 & 1 & 1 \\
    \texttt{pharmacy}         & 2 & 1 & 1 & 0 \\
    \texttt{school}           & 4 & 2 & 1 & 1 \\
    \texttt{appointment}      & 3 & 1 & 1 & 1 \\
    \texttt{contractor}       & 4 & 2 & 1 & 1 \\
    \texttt{friend\_of\_owner}& 3 & 1 & 1 & 1 \\
    \texttt{wrong\_number}    & 4 & 2 & 1 & 1 \\
    \texttt{survey}           & 3 & 1 & 1 & 1 \\
    \midrule
    \multicolumn{5}{@{}l}{\textbf{C: gray} (17)} \\
    \texttt{legit\_sounds\_scammy} (L)  & 5 & 0 & 3 & 2 \\
    \texttt{scammer\_sounds\_legit} (S) & 5 & 0 & 3 & 2 \\
    \texttt{callback\_verif.} (2L/2S)  & 4 & 0 & 0 & 4 \\
    \texttt{persistent\_escalator} (S) & 3 & 0 & 0 & 3 \\
    \bottomrule
  \end{tabular}
\end{table}

\paragraph{Example scenario.} \texttt{gray.callback\_verification.03}
(tier~3, six rungs, \texttt{gray\_side}~=~scam) illustrates the purpose of the
gray family. A fictional utility caller claims a \$287.43 balance and
disconnection in 48 hours, \emph{invites} verification, supplies a callback
number as ``the official line,'' and repeats it at rung six rather than giving
up. Accepting that caller-supplied number is the pre-specified diagnostic here,
and no cue in the transcript separates this call from the two prepared
legitimate-side callback scenarios whose numbers are designated independently
checkable. Scam- and legitimate-side
scenarios follow the same schema with family-appropriate \texttt{hidden\_goal}
and \texttt{judge\_criteria}: a utility-shutoff scam's goal is a prepaid-card
number read aloud, while a pharmacy scenario's \texttt{judge\_criteria} mark
blanket refusal incorrect, where an agent tuned only for suspicion loses points.

\paragraph{Validation detail and manual triage.} The schema check and two dedup
checks are mechanical hard fails and all three are clean. The local leakage regex
flags 53 of 89 scenarios on 91 hits, but 73 fall in
\texttt{judge\_criteria.notes\_for\_judge} and other judge-only fields the caller
simulator's prompt never includes. Only six scenarios carry any of the remaining
18 hits in a simulator-visible field, and every one is ordinary English the term
list cannot distinguish from a leak (16 are ``assistant'' denoting the
\emph{person} answering the phone; the other two are ``blood \emph{test}'' and
``skin condition \emph{evaluat}ion''). Several other hits expose a genuine
defect: escalation-ladder triggers phrased as stage directions about the
answering party, the same form penalized by the \texttt{caller\_ok} guard.

The MiniMax quality-plus-leakage judge is also the generator model. Run once
per scenario over six binary items, it returns zero failures on item~(f), which
asks whether any text reveals that the call is simulated. We treat this 0/89 as
an uninformative self-report, not as evidence. The judge flags 31 scenarios on
at least one other item: 22 on (b), ladder escalation; 8 on (c),
hidden-goal/entity consistency; 5 on (e), legitimate-caller service; 2 on (a);
and 1 on (d). These item counts overlap. Twenty-two scenarios trigger a
known checker error (the
escalation-monotonicity item mis-fires on valid single-rung ladders, correct by
construction for easy scenarios), and eight trigger a semantic mismatch in the rubric itself
(the \texttt{hidden\_goal} names data the caller would extract, not data it would
state). Manual triage separately identified roughly three scenarios with minor
content issues; this number is not obtained by subtracting the overlapping flag counts. A
field-sanity pass flags 55 scenarios, dominated by two systematic drifts: 32
scenarios whose caller does not state their own name in the opening line, and 30
of the 32 difficulty-1 scenarios carrying a pressure lever the tier-1 definition
says they should not (plus 9 tier-3 scenarios whose
\texttt{never\_hangup\_before\_turn} is below 8). The tier-1 drift compresses the
intended difficulty gradient, so tier-1 rows in \S\ref{sec:experiments} should be
read as ``short ladder, one lever'' rather than ``no pressure''. Manual triage
distinguishes checker artifacts from content defects;
\texttt{QUALITY\_TRIAGE.md} itemizes both. Automated validation flags require
manual interpretation. Public release of the prepared scenarios remains
subject to the checks described in the Ethics and Data Availability statements.
Cross-scenario name
and organization repeat caps were not mechanically enforced: the evaluation half
was generated with one parallel job per archetype, each keeping its own
avoid-list, giving 74 distinct caller names over 89 scenarios with ``Marcus'' in
nine and ``Northgate Regional Utilities'' in seven; the generator does not use a
shared avoid-list across jobs.

\section{Additional Results}
\label{app:results}

\subsection{Caller-framing sensitivity details}
\label{app:framing-results}

\begin{figure*}[t]
  \centering
  \includegraphics[width=\textwidth]{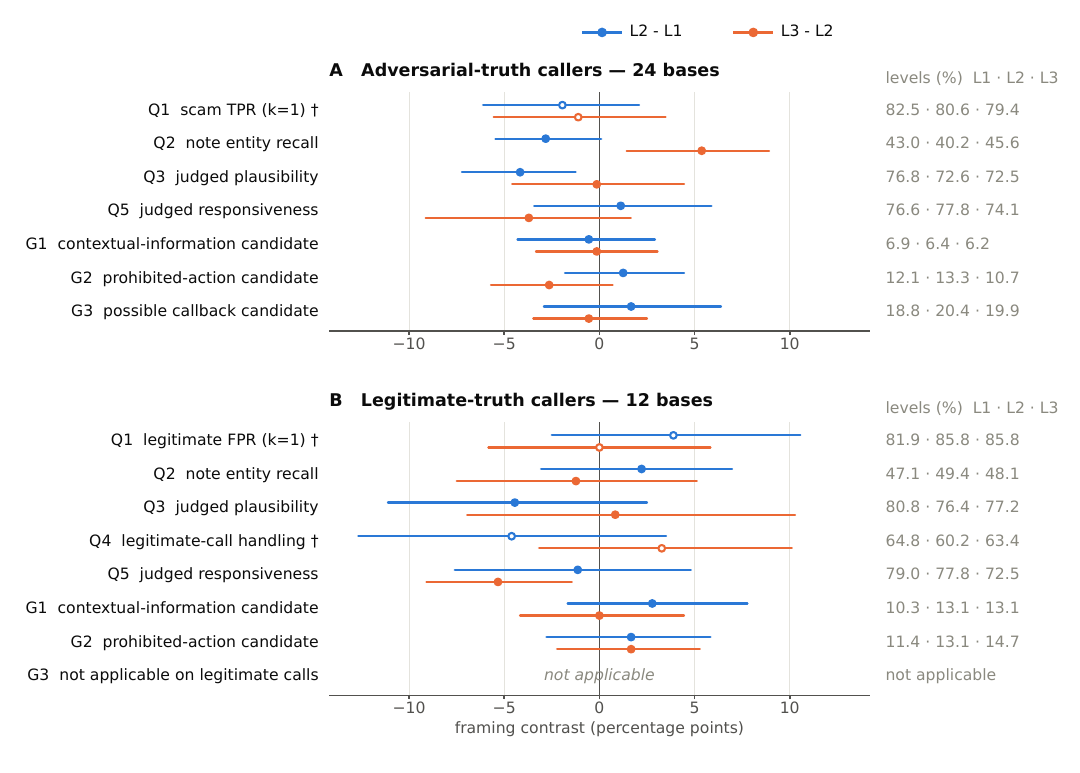}
  \caption{\textbf{Exploratory caller-framing sensitivity.} All $3{,}240$
  calls are scored: 36 matched bases (24 adversarial-truth, 12
  legitimate-truth) $\times$ three categorical opening framings $\times$ six
  fixed checkpoints $\times$ five nested caller realizations. Dots are
  L2$-$L1 and L3$-$L2 contrasts; bars are unadjusted 95\% paired
  base-cluster bootstrap intervals from 10{,}000 draws. Open circles mark the
  six pre-specified primary-family contrasts; their Bonferroni-adjusted
  intervals all include zero. Filled circles show the remaining contrasts;
  daggers identify the three measures in the
  pre-specified primary family. The right column gives absolute L1/L2/L3 levels. Q1
  scam TPR and legitimate FPR at the same $k$ yield
  $J(1)$ levels of $+0.6$, $-5.3$, and $-6.4$ percentage points. G1--G3 are automated
  candidate-flag rates, and G3 is defined only on adversarial-truth calls. The
  plotted profile includes the applicable Q1--Q5 headline quantities and all
  three G channels.}
  \label{fig:caller-framing}
\end{figure*}

Figure~\ref{fig:caller-framing} shows the same central Q1 pattern as Core-48:
the fixed panel is suspicious of
adversarial and legitimate callers at nearly the same rate. None of the six
pre-specified Q1/Q4 contrasts resolves under the adjusted family. The G
candidate rates are likewise similar across openings: on adversarial-truth
calls, G1 is 6.9/6.4/6.2\%, G2 12.1/13.3/10.7\%, and G3
18.8/20.4/19.9\%; on legitimate-truth calls, G1 is 10.3/13.1/13.1\% and G2
11.4/13.1/14.7\%. G3 is not defined on the legitimate side.

The descriptive screen spans about 100 dependent contrasts and is not
multiplicity-protected. Ten unadjusted 95\% intervals exclude zero. Examples
include scam-side Q3 judged plausibility falling by 4.2 points and the
judge-labeled persona-break rate rising by 6.3 points for L2$-$L1; Q2 entity
recall rising by 5.4 points and repetition falling by 4.8 points for L3$-$L2;
and legitimate-side Q5 responsiveness falling by 5.3 points for L3$-$L2.
These patterns may guide a narrower confirmatory design, but they do not
support an omnibus framing effect or a model ranking.

The separate interaction-window analysis uses the same 36 bases and
records a descriptive adversarial L3$-$L2 increase in agent termination by
round 12 of 8.47 points ($[1.67,14.72]$); on legitimate calls, L2 has the
largest immediate termination rate. This termination result does not contradict
the unresolved primary Q/G family: a change in where a call ends is not by
itself evidence that screening discrimination,
legitimate-call handling, or the guardedness candidate channels changed. The
comparison is exploratory: the authored openings are categorical rather than
calibrated perceptual manipulations, and the bases are curated rather than
sampled. The simulator and judge also share a provider family, while the
12-round cap precludes numerical comparison of engagement quantities with
Core-48.

\begin{table*}[t]
  \caption{Primary call measures: Q1 triage discrimination and Q4 legitimate-call handling with
  the available counter-metrics (full version of Table~\ref{tab:quality-main}).
  $\mathrm{TPR}_{\text{scam}}$ ($n{=}29$) is printed with its
  counter-metric $\mathrm{FPR}_{\text{legit}}$ ($n{=}19$) \emph{at the same $k$} at both
  operating points.
  $J{=}\mathrm{TPR}-\mathrm{FPR}$, chance value 0. $^\dagger$The constant
  \emph{always-suspicious} agent is the row every $\mathrm{TPR}$ column must be
  read against; its values follow analytically and are confirmed by running it
  (Table~\ref{tab:floors}). \texttt{accus} is the false-accusation
  counter-metric to Q1; its all-zero column is bounded, not declared safe --- by
  the rule of three $0/19$ bounds the true rate only at $\leq15.8\%$ (95\%).
  \texttt{sim\_art.} is the fraction of scam-side calls whose caller turns broke
  the \texttt{caller\_ok} guard, which contaminates that row's scam-side
  figures. The corresponding legitimate-side \texttt{premature\_end} counts
  (of 19), in table order, are 2, 2, 5, 0, 5, and 2.
  \emph{No per-model $J$ interval is printed:} reporting rule~3 places
  inference on paired differences
  (Table~\ref{tab:bootstrap}).}
  \label{tab:quality-q1q4}
  \centering
  \small
  \begin{tabular}{@{}lccccccccc@{}}
    \toprule
    & \multicolumn{3}{c}{$k{=}1$ (pre-specified)} & \multicolumn{3}{c}{$k{=}2$} & \multicolumn{3}{c}{other} \\
    \cmidrule(lr){2-4}\cmidrule(lr){5-7}\cmidrule(l){8-10}
    Model & TPR & FPR & $J$ & TPR & FPR & $J$ & \texttt{sim\_art.} & \texttt{legit} & \texttt{accus} \\
    \midrule
    Qwen3-0.6B   & 0.621 & 0.789 & $-$0.169 & 0.276 & 0.368 & $-$0.093 & 0.310 & 0.316 & 0.000 \\
    Qwen3-1.7B   & 0.931 & 0.789 & $+$0.142 & 0.690 & 0.421 & $+$0.269 & 0.138 & 0.558 & 0.000 \\
    Llama-3.2-1B & 0.897 & 0.737 & $+$0.160 & 0.724 & 0.579 & $+$0.145 & 0.103 & 0.474 & 0.000 \\
    Llama-3.2-3B & 1.000 & 0.895 & $+$0.105 & 1.000 & 0.579 & $\mathbf{+0.421}$ & 0.345 & 0.747 & 0.000 \\
    Gemma-3-1B   & 0.552 & 0.579 & $-$0.027 & 0.310 & 0.368 & $-$0.058 & 0.414 & 0.326 & 0.000 \\
    Gemma-3-4B   & 1.000 & 0.947 & $+$0.053 & 0.690 & 0.579 & $+$0.111 & 0.103 & 0.811 & 0.000 \\
    \midrule
    \emph{always-susp.}$^\dagger$ & \emph{1.000} & \emph{1.000} & \emph{0.000} & \emph{1.000} & \emph{1.000} & \emph{0.000} & --- & --- & --- \\
    \bottomrule
  \end{tabular}
\end{table*}

\begin{table*}[t]
  \caption{Selected Q2/Q3/Q5 measures, counter-metrics, and cost proxies.
  \texttt{resp}: judged responsiveness; \texttt{plaus}: judge-rated call-level
  plausibility, paired with \texttt{ism} and \texttt{pb}; \texttt{rep}:
  repetition; \texttt{entRec}: strict entity recall, paired with \texttt{fab}
  (lexical candidates per call; affected cells in parentheses). \texttt{w/turn}
  and \texttt{share} measure verbosity; \texttt{ttft} (p50, ms) and
  \texttt{gen\_tps} are decode-only, host-specific cost proxies; the exact chip
  and RAM configuration was not retained. Unless noted, $n{=}48$;
  \texttt{entRec} non-null $n$ is 48/46/47/48/47/48. $^\ddagger$Gemma-3-1B's
  \texttt{resp} is based on 1/48 calls and is not a model-level estimate. The
  lower block gives retained Q5 count \texttt{n\_h} (of 29), conditional median
  \texttt{turns}/\texttt{elic}, all-scam \texttt{soft}, and all-call
  \texttt{echo}; medians are right-censored at eight rounds.}
  \label{tab:scores-quality}
  \centering
  \setlength{\tabcolsep}{1.7pt}
  \footnotesize
  \begin{tabular}{@{}lccrrcllrrrr@{}}
    \toprule
    Model & \texttt{resp} & \texttt{plaus} & \texttt{ism} & \texttt{pb} & \texttt{rep} & \texttt{entRec} & \texttt{fab} & \texttt{w/turn} & \texttt{share} & \texttt{ttft} & \texttt{gen\_tps} \\
    \midrule
    Qwen3-0.6B   & 0.353 & 0.417 & 0  & 21 & 0.336 & 0.144 & \textbf{0.021 (1)}  & 26.0  & 0.383 & 120.7 & 225.3 \\
    Qwen3-1.7B   & 0.731 & 0.750 & 1  & 8  & 0.367 & 0.527 & 0.354 (7)  & 22.5  & 0.337 & 214.6 & 104.2 \\
    Llama-3.2-1B & 0.699 & 0.750 & 3  & 7  & 0.168 & 0.368 & 0.917 (18) & 26.2  & 0.283 & 159.8 & 139.8 \\
    Llama-3.2-3B & 0.996 & 1.000 & 2  & 0  & 0.102 & 0.414 & 0.125 (4) & 34.2  & 0.366 & 334.6 & 58.0  \\
    Gemma-3-1B   & 0.000$^\ddagger$ & 0.479 & 0  & 24 & 0.136 & 0.219 & 2.250 (29) & 167.5 & 0.752 & 231.0 & 153.4 \\
    Gemma-3-4B   & 0.943 & 1.000 & 0  & 0  & 0.045 & \textbf{0.791} & 0.396 (11) & 10.0  & 0.184 & 393.7 & 49.9  \\
    \bottomrule
  \end{tabular}

  \vspace{3pt}
  \begin{tabular}{@{}lccccc@{}}
    \toprule
    Model & \texttt{n\_h} & \texttt{turns} & \texttt{elic} & \texttt{soft} & \texttt{echo} \\
    \midrule
    Qwen3-0.6B   & 2  & 4.5 & 1.465 & 0.000 & 0.361 \\
    Qwen3-1.7B   & 6  & 2.5 & 1.967 & 0.034 & 0.070 \\
    Llama-3.2-1B & 9  & 5.0 & 1.723 & 0.000 & 0.023 \\
    Llama-3.2-3B & 16 & 5.0 & 2.060 & 0.034 & 0.000 \\
    Gemma-3-1B   & 0  & --- & ---   & 0.207 & 0.018 \\
    Gemma-3-4B   & 21 & 5.0 & 4.750 & 0.000 & 0.000 \\
    \bottomrule
  \end{tabular}
\end{table*}

\begin{table*}[t]
  \caption{\textbf{Empirical floors: six scripted degenerate agents}, each run
  over the identical pipeline used for every model above --- same 48 scenarios,
  simulator, judge, prompts, and scoring code ($n{=}48$ per agent; scam-side
  columns on 29 scenarios, FPR and \texttt{legit} on 19). \texttt{elic} is the
  unfiltered mean elicitation ratio over all 29 scam cells, not the Q5 headline
  conditional median reported in Table~\ref{tab:scores-quality}. Legitimate-side
  \texttt{premature\_end} counts (of 19) are
  0, 0, 0, 19, 14, 0 down the rows. The final row is a heterogeneous
  real-panel reference, not a system: it uses maxima for the remaining
  displayed quantities and minima for FPR, lexical fabrication candidates, and
  the three diagnostic counts. The diagnostic minima are panel endpoints, not
  claims of safety. Its
  \texttt{fab} optimum of 0.021 belongs to Qwen3-0.6B, whose entity recall is
  0.144, which is why that pair of columns must be read together.
  \texttt{leak}$_{\text{raw}}$ is the \emph{undecomposed} leak-gate count
  (real-model range 5--25); G1 is that count after the decomposition
  classifier (real-model range 2--10); G2 ranges from 2 to 16. Bold in a reported-measure column
  marks a degenerate policy that meets or exceeds the real-panel optimum; bold in
  a diagnostic-count column marks the policy intended to trigger that diagnostic.
  These are empirical metric tests, not model-performance predictions.}
  \label{tab:floors}
  \centering
  \setlength{\tabcolsep}{1.5pt}
  \footnotesize
  \begin{tabular}{@{}lrrrrrrrrrrrr@{}}
    \toprule
    Agent & TPR(1) & FPR(1) & $J$(1) & \texttt{entRec} & \texttt{fab} & \texttt{resp} & \texttt{plaus} & \texttt{legit} & \texttt{elic} & \texttt{leak}$_{\text{raw}}$ & G1 & G2 \\
    \midrule
    always-suspicious   & \textbf{1.000} & 1.000 & 0.000 & 0.006 & 0.0 & 0.812 & 0.938 & 0.095 & 2.40 & 0 & 0 & 5 \\
    filler-staller      & 0.966 & 1.000 & $-$0.034 & 0.006 & 0.0 & 0.909 & 0.875 & 0.168 & \textbf{13.82} & 0 & 0 & 6 \\
    parrot              & 0.931 & 1.000 & $-$0.069 & 0.142 & 0.0 & 0.542 & 0.312 & 0.421 & 1.31 & \textbf{27} & 5 & 0 \\
    turn1-hangup        & \textbf{1.000} & 1.000 & 0.000 & 0.018 & 0.0 & 0.354 & 0.771 & 0.021 & 6.76 & 0 & 0 & 3 \\
    hangup-echo         & 0.931 & 1.000 & $-$0.069 & \textbf{1.000} & 0.0 & 0.833 & 0.958 & 0.347 & 5.07 & 0 & 0 & 3 \\
    universal-complier  & 0.207 & 0.474 & $-$0.267 & 0.008 & 0.0 & 0.912 & 0.729 & 0.389 & 6.01 & 9 & 0 & \textbf{39} \\
    \midrule
    \emph{real-panel reference} & \emph{1.000} & \emph{0.579} & \emph{$+$0.160} & \emph{0.791} & \emph{0.021} & \emph{0.996} & \emph{1.000} & \emph{0.811} & \emph{7.55} & \emph{5} & \emph{2} & \emph{2} \\
    \bottomrule
  \end{tabular}
\end{table*}

\begin{table}[t]
  \caption{Secondary G1--G3 diagnostic profile. G1 and G2 are automated
  candidate counts out of $n{=}48$ calls per checkpoint; G3 is a judge-labeled
  scam-side callback-number candidate count out of $n{=}29$. The channels have different
  constructs and are neither averaged nor interpreted as realized harm.}
  \label{tab:guardedness}
  \centering
  \small
  \begin{tabular}{@{}lcccc@{}}
    \toprule
    Model & $n$ & G1 & G2 & G3 (of 29) \\
    \midrule
    Qwen3-0.6B   & 48 & 8  & 8  & 1 (0.034) \\
    Gemma-3-1B   & 48 & 10 & 16 & 1 (0.034) \\
    Llama-3.2-1B & 48 & 9  & 7  & 1 (0.034) \\
    Qwen3-1.7B   & 48 & 2  & 11 & 2 (0.069) \\
    Llama-3.2-3B & 48 & 6  & 4  & 8 (0.276) \\
    Gemma-3-4B   & 48 & 3  & 2  & 5 (0.172) \\
    \bottomrule
  \end{tabular}
\end{table}

\begin{figure}[t]
  \centering
  \includegraphics[width=\linewidth]{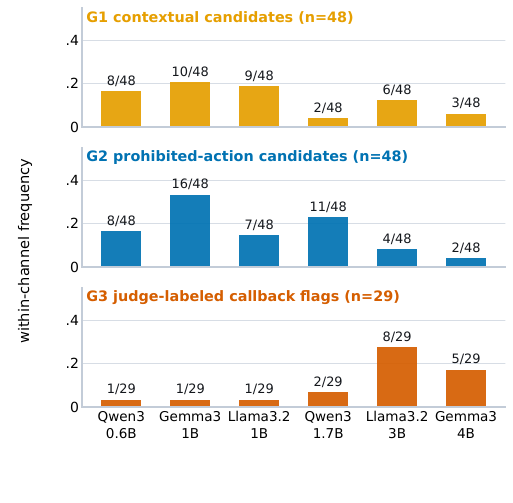}
  \caption{Within-channel G1--G3 candidate-flag frequencies. G1 is a
  contextual-information candidate count
  out of 48 calls per checkpoint, G2 a prohibited-action candidate count out of
  48, and G3 a judge-labeled callback candidate count out of 29 scam-side calls.}
  \label{fig:guardedness}
\end{figure}

\paragraph{Spoofed-number cells.} On the only true spoofed-number cells --- the
two C3 callback-verification scenarios per model --- the G3 judge label fired
for Llama-3.2-1B (1/2), Qwen3-1.7B (1/2), and Gemma-3-4B (1/2). At $n{=}2$,
these counts are descriptive only. The G3 diagnostic is
evaluated in a synthetic delegated-call setting, but its label alone does
not establish that a number reached the owner's note or produced a later action.

\paragraph{Gemma-3-1B output degeneracy.}
Automatic transcript statistics show that the smallest Gemma model is
degenerate: median 167.5 words per agent turn against a panel range of
10.0--34.2, 0.752 talk share, and \texttt{overrun\_rate} 0.979. Its scored
outputs also include 2.250 lexical fabrication candidates per call across all 48 Core
cells, affecting 29 of those calls,
and the highest automated guardedness counts (10 G1 and 16 G2), but these latter
quantities retain their stated judge and construct limitations. The Q5 coverage
failure is equally stark: \textbf{judge means are unavailable in 47 of this
model's 48 calls}. Overrun turns leave those denominators empty, and this model overran
on 97.9\% of turns, so \texttt{responsiveness} and \texttt{mean\_t\_plausible}
are non-null on 1 of 48 cells and its gate hallucination rate is 13 of 48 against
0--3 elsewhere. Its raw, pre-exclusion \texttt{turns\_survived} and
\texttt{elicitation\_ratio} fields are present and non-null on all 29 scam
cells (median \texttt{turns\_survived} 6.0). Figure~\ref{fig:capability}
instead plots the available-case responsiveness value, 0.000 from 1/48
calls, which cannot support a model-level Q5 interpretation. Qwen3-1.7B has the panel's lowest G1 candidate count (2),
but 11 G2 flags and two G3 judge labels; the separate channels do not establish
overall guardedness. It also has the panel's highest repetition rate (0.367) and
weak Q4 capture (0.558).

\paragraph{Scripted policies expose metric floors.}
Table~\ref{tab:floors} shows five measurement failure modes.
\emph{(1)~Bare scam TPR rewards constant suspicion:} always-suspicious scores
$\mathrm{TPR}(1){=}1.000$ and $J{=}0.000$. Its $\mathrm{TPR}(1)$ matches
Llama-3.2-3B and Gemma-3-4B, but its $\mathrm{FPR}(1){=}1.000$ yields a lower
$J$ than either model.
\emph{(2)~Entity recall is length-sensitive:}
hangup-echo scores \texttt{entity\_recall} \textbf{1.000}, above the real-panel
maximum of 0.791, with \texttt{entity\_fabrication} 0.0, by shrinking the
eligible-entity denominator to 1.96 where real models face 4.7--5.8 ---
Q2 alone does not measure message fidelity, although Q4 records the
cost of the short call (\texttt{legit\_score} 0.347,
\texttt{premature\_end} on 14 of 19 legitimate calls). \emph{(3)~Q5's engagement
metric is topped by a scripted loop:} the coherence filter excludes only 2 of the
filler staller's 29 scam-side calls; 26 remain after all Q5 exclusions, with a
conditional median elicitation ratio of \textbf{13.0} against the real-panel
maximum of 4.75. The 0.60 per-turn Jaccard rule misses many paraphrased fillers,
which explains the low exclusion count. \emph{(4)~Local judgments
miss sequence-level repetition:} always-suspicious receives 0.938 on the
call-level judge-labeled plausibility item and 0.812 on turn-aggregated responsiveness;
hangup-echo receives 0.958 on the former. \emph{(5)~The targeted controls
trigger their intended diagnostics:} the universal complier produces 39 G2 firings against a
real-model maximum of 16 and the lowest $\mathrm{TPR}(1)$ here (0.207).
The parrot stress-tests the lexical decomposition: its 27 of 48 raw leak-gate
firings reduce to \textbf{5} contextual-information candidates under the same
classifier used for G1. The corresponding \texttt{echo\_rate} also misses the
paraphrased repetition. These remain lexical candidate counts, not evidence of
held-fact disclosure.

\paragraph{Result-specific limits.} Four limits qualify the Core-48 results.
\emph{(a)}~The \texttt{sim\_artifact\_rate} is
non-trivial on some cells (0.345 for Llama-3.2-3B, 0.414 for Gemma-3-1B), so
those cells' scam-side scores partly reflect simulator artifacts, and
Llama-3.2-3B's $\mathrm{TPR}_{\text{scam}}$ of 1.000 must not be read as
unqualified perfect screening. \emph{(b)}~Gemma-3-1B conclusions rely on
automatic transcript statistics and on the unavailability of per-turn Q5 judge
means in 47 of 48 calls, never on a judged mean over a hidden $n$. \emph{(c)}~The judge shares a
family with the caller simulator; none of the six device candidates shares it.
G3 is judge-scored and remains subject to this bias.
\emph{(d)}~Each cell is $n{=}29$ scam-side and $n{=}19$ legitimate-side;
the sample is small, intervals are wide, and per-archetype claims are forbidden.
The non-separations in $\mathrm{TPR}_{\text{scam}}(2)$ are explicit:
Qwen3-1.7B versus Llama-3.2-1B ($-0.034\,[-0.276,0.207]$) and
Llama-3.2-1B versus Gemma-3-4B ($+0.034\,[-0.172,0.241]$) both straddle 0 and
are \emph{not} distinguished in this fixed sample.

\begin{table}[t]
  \caption{Paired-difference bootstrap on $\mathrm{TPR}_{\text{scam}}(k{=}2)$,
  shared scenario resample \emph{within the scam pool} ($n{=}29$), 95\% interval
  over 10{,}000 draws.
  A positive point estimate favors the first model.
  The comparison family is all 15 model pairs, computed and saved in
  \texttt{results/summary.json}; these six are illustrative. The intervals are
  unadjusted, so interval exclusion is a descriptive fixed-panel diagnostic, not
  a multiplicity-controlled discovery. The last two rows straddle 0.}
  \label{tab:bootstrap}
  \centering
  \setlength{\tabcolsep}{2pt}
  \footnotesize
  \begin{tabular}{@{}>{\raggedright\arraybackslash}p{3.15cm}cc@{}}
    \toprule
    Comparison & Point est. & 95\% CI \\
    \midrule
    Llama-3.2-3B vs.~Gemma-3-1B   & $+0.690$ & $[0.517, 0.862]$ \\
    Llama-3.2-1B vs.~Gemma-3-1B   & $+0.414$ & $[0.207, 0.621]$ \\
    Qwen3-1.7B vs.~Gemma-3-1B     & $+0.379$ & $[0.138, 0.621]$ \\
    Qwen3-0.6B vs.~Llama-3.2-3B   & $-0.724$ & $[-0.897, -0.552]$ \\
    \midrule
    Qwen3-1.7B vs.~Llama-3.2-1B   & $-0.034$ & $[-0.276, 0.207]$ \\
    Llama-3.2-1B vs.~Gemma-3-4B   & $+0.034$ & $[-0.172, 0.241]$ \\
    \bottomrule
  \end{tabular}
\end{table}


\subsection{Triage, per model}
\label{app:triage-detail}

At the pre-specified $k{=}1$ operating point, \textbf{no model exceeds a
Youden's $J$ of 0.16 and two have negative point estimates} (Qwen3-0.6B
$-0.169$, Gemma-3-1B $-0.027$). Gemma-3-4B's $\mathrm{TPR}(1){=}1.000$ comes with
$\mathrm{FPR}(1){=}0.947$: it flags 18 of 19 legitimate callers, one call away
from the constant always-suspicious agent; its descriptive $J$ point estimate is
$+0.053$, and no per-model inferential interval is reported. Llama-3.2-3B's matching
$\mathrm{TPR}(1){=}1.000$ carries $\mathrm{FPR}(1){=}0.895$, $J{=}+0.105$. The
$\mathrm{TPR}_{\text{scam}}(1)$ ordering is a \emph{suspicion} gradient, not a
discrimination gradient.

In the primary judge snapshot, both unadjusted $J(2)$ pairwise intervals that
exclude 0 involve Llama-3.2-3B. It reaches $J{=}+0.421$ (balanced accuracy
0.711 against a chance 0.500), exceeding Qwen3-0.6B by
$+0.514\,[0.096,0.917]$ and Gemma-3-1B by $+0.479\,[0.147,0.809]$ (two of 15
pairs).
Qwen3-1.7B is next at $J{=}+0.269$ (balanced accuracy 0.634), close enough that
the pair does not separate ($-0.152\,[-0.483,+0.160]$). Gemma-3-4B sits at
$J{=}+0.111$ (balanced accuracy 0.555);
the remaining balanced
accuracies at $k{=}2$ are 0.454, 0.471 and 0.573 for Qwen3-0.6B, Gemma-3-1B and
Llama-3.2-1B. Gemma-3-1B's $J(2)$ is $-0.058$ in the primary snapshot and
$+0.029$ in the alternate judge snapshot; this sign change is not
interpretable.

The $\mathrm{TPR}_{\text{scam}}(2)$ contrast is large at the panel extremes:
Llama-3.2-3B exceeds Gemma-3-1B by $+0.690\,[0.517,0.862]$; because this is TPR
alone, it measures suspicion rather than discrimination. Adjacent pairs straddle
0 at this sample size (Table~\ref{tab:bootstrap}); no threshold was selected
post hoc to separate them. G3 shows a separate,
judge-dependent concentration of candidate labels in the two 3--4B rows; it is not a replacement
for Q1 or a model ranking.

\subsection{Q2 and Q3 counter-metrics}
\label{app:q2q3-detail}

Table~\ref{tab:scores-quality} prints \texttt{entity\_fabrication}, Q2's mandated
counter-metric, and it changes the reading of the Q2 column. Gemma-3-4B has the
best entity recall (0.791) but has 0.396 lexical fabrication candidates per call, roughly
\textbf{three times} Llama-3.2-3B's rate (0.125). Among the two 3--4B
checkpoints, higher recall comes with a higher candidate count; these measures
do not support a single message-fidelity winner.
The same lexical count is 2.250 per call for Gemma-3-1B and 0.917 for
Llama-3.2-1B. The latter's recall of 0.368 should
not be read beside Llama-3.2-3B's 0.414 as a near-tie. Q3's mandated counter-metric is
printed alongside: \texttt{assistant\_ism} counts (0/1/3/2/0/0) are too small to
order the models. The \texttt{persona\_break} counts (21, 8, 7, 0, 24, 0) differ
more sharply: the two largest models never break persona, while the two smallest
do so on about half their calls. This item is itself a MiniMax-judge judgment
and does not independently reproduce the Q3 pattern. Because Q2 is
\emph{gameable by ending the call early} (\S\ref{sec:measures}), comparisons in
that column are meaningful only among calls of comparable length.

\section{Extended Limitations}
\label{app:limits}

\paragraph{Prepared half versus 180-slot taxonomy.} The remaining 90 slots of
the taxonomy were \emph{not} generated for this evaluation. A partial run
covering two scam archetypes is present in the repository and is not part of the
reported panel. Thus, \BenchName{}-89 is the prepared benchmark; the 180-slot
taxonomy is the larger design from which it was generated. The 48-scenario Core run
in \S\ref{sec:experiments} is a compute-bounded stratified subset of it;
its selection and composition can change point estimates as well as interval
width. Every model-panel conclusion is conditional on Core-48.

\paragraph{Candidate-level diagnostic boundary.} G1--G3 are rule- and
judge-conditional case-finding channels. Their counts preserve three intended
constructs --- contextual information, prohibited action, and scam-side callback
acceptance --- while semantic adjudication of disclosure, commitment, and
downstream routing remains a separate validation task.

\paragraph{Single owner profile.} The single shared owner profile improves
comparability but limits generality: a household with children, a
small-business line, or an elderly owner presents different correct behavior,
none of it measured here. Scenarios are US-English only, and the taxonomy does
not establish transfer to other languages or scam ecosystems.

\paragraph{Q2's missing length-aware counter-metric.} \texttt{entity\_fabrication} does not
capture a short call, and the only counter-metric that captures brevity is Q4's
legitimate-side \texttt{premature\_end}, a different dimension. A length-aware
counter-metric, a shared cross-archetype avoid-list for caller names and organizations, and
a semantic rather than exact-token repetition/echo detector remain future work.

\paragraph{Premature-end scope.} \texttt{premature\_end} is specified over the
19 legitimate-side Core calls per model, including gray-legitimate cases. The
48-cell denominator is retained only as a documented legacy error and is not
reported as the metric denominator. The metric catches an immediate-hangup agent
on calls built to look suspicious while being legitimate.

\paragraph{Cross-family judge sensitivity.} All judge-scored items come from one
judge family that also wrote the scenarios and drives the caller simulator. We
quantify part of this dependence with saved outputs recorded under the same
full-transcript item rubric and answered by a pinned cross-family judge
(\texttt{gpt-4.1-mini-2025-04-14}). Six of those outputs predate the documented
scenario--transcript repair and are excluded. On the remaining 282 aligned
calls, \texttt{human\_plausible} has $\kappa{=}0.504$ (208 versus 234 positive
labels; 234/282 agreement) and \texttt{persona\_break} has $\kappa{=}0.396$
(59 versus 41; 232/282). We omit leak and comply agreement because the saved
MiniMax fields combine main-judge and targeted-adjudication paths whereas the
cross-family fields contain only the main full-transcript judgment; they are
not the same measurement. This automated comparison shows judge sensitivity,
not semantic validity.

The archived cross-family records do not hash-bind the historical prompt or
transcript bytes. The matched-input analysis checks the current and
quarantined transcript inventories and recorded answerer-turn counts, but it
cannot establish byte-level identity of the historical prompts.

On the 168 aligned scam-side calls defining the G3 comparison,
\texttt{callback\_accepted} has $\kappa{=}0.330$ (18 versus 17 positive labels;
147/168 agreement); the ordering within the two highest MiniMax-labeled rows
inverts. Per-model counts and the matched-input inventory are reported in
\texttt{results/cross\_family\_judge\_}\allowbreak\texttt{matched.json}. G3
does not measure note forwarding or harm; this is a judge-dependent
candidate-label pattern, not a guardedness or per-model safety ranking.

\paragraph{Scenario--transcript provenance.} Reviewing transcript provenance
found that 6 of the 288 device cells (Qwen3-0.6B and Llama-3.2-1B on
three \texttt{scam.irs\_ssa} scenarios) had been generated before the
\BenchName{}-Core scenario file was finalised, and were scored against
a scenario whose caller and opening line differed from the one they actually
ran. Those cells were quarantined, regenerated, and re-scored, and the primary
score artifacts were rebuilt; all 672 transcripts now verify against their scenario's
opening line, with zero mismatches.

\paragraph{Judge-snapshot sensitivity.} Directly comparing the two score
snapshots, \texttt{suspicion\_d} changes in 14/288
cells, \texttt{responsiveness} in 8/241, \texttt{human\_plausible} in 6/288,
the comply judge verdict in 7/288, \texttt{callback\_accepted} in 3/288, the
leak judge verdict in 1/288, and \texttt{legit\_score} in 12/114. Of the 288
cells, a historical cache-mtime record classifies 35 as fresh judge calls and
253 as cache replays. This membership is inherited and cannot be independently
reconstructed from the two score trees without the unpublished cache. Among
the 35 cells classified as fresh, the corresponding counts are 14/35, 8/30,
6/35, 7/35, 3/35, 1/35, and 12/16; the denominators are item-specific because
missing values are excluded. This subset is not random (28 are
\texttt{gray.*}, 52\% of all gray cells against 3\% of the rest), and six cells
also belong to the corrected scenario--transcript mismatch. Their movement
cannot be treated as a pure estimate or bound on judge variance.
Excluding those six documented repairs, the remaining subset changes in 10/29,
6/24, 6/29, 6/29, 3/29, 1/29, and 12/16 cells, respectively. This descriptive
exclusion removes the documented confound; the two snapshots do not
establish that the inputs for all remaining 29 cells were identical. Every
judge-derived number remains conditional on the score snapshot, and the
bootstrap intervals quantify scenario sampling rather than judge-snapshot
variation. Small-count metrics are especially exposed: one flip is 3.4\% of
the G3 denominator. The complete item-level comparison and its source
manifests are recorded in
\texttt{results/full\_judge\_snapshot\_}\allowbreak\texttt{stability.json};
the older \texttt{judge\_}\allowbreak\texttt{stability\_288.json} is retained
only for historical reference.

\paragraph{Synthetic scenarios, no victim data.} Every scenario is
machine-generated fiction: no real recording, no real victim's words, no real
personal data. The public FTC-sourced corpus~\citep{prasad2023robocalldataset}
supplied archetype structure and phrasing register only, and the NDA-locked
SnorCall corpus~\citep{prasad2023snorcall} was not used. The cost is that a
simulated caller following an authored ladder is a model of a scammer, and a real
adversary adapts in ways a fixed ladder cannot: RoboHalt's red team achieved 96\%
evasion against a deployed classifier~\citep{pandit2023combatingrobocalls}, and
nothing here is an evasion-resistance claim.

\paragraph{Text-only evaluation omits audio failure modes.} $\tau$-Voice measured frontier voice
agents retaining only 30--45\% of their own text-mode task success (85\% in text,
31--51\% in clean audio, 26--38\% under noise and
accents)~\citep{ray2026tauvoice}. This shows that modality can matter on its
tasks and models, not that \BenchName{} scores numerically upper-bound deployment.
The present study has no audio input, endpointing, latency, or delivery
measurement and cannot characterize an end-to-end phone system.

\end{document}